# Quantitative Molecular Scaling Theory of Protein Amino Acid Sequences, Structure, and Functionality

J. C. Phillips

Dept. of Physics and Astronomy, Rutgers University, Piscataway, N. J., 08854

Abstract

Here we review the development of protein scaling theory, starting from backgrounds in mathematics and statistical mechanics, and leading to biomedical applications. Evolution has organized each protein family in different ways, but scaling theory is both simple and effective in providing readily transferable dynamical insights complementary for many proteins represented in the 90 thousand static structures contained in the online Protein Data Base (PDB). Scaling theory is a simplifying magic wand that enables one to search the hundreds of millions of protein articles in the Web of Science, and identify those proteins that present new cost-effective methods for early detection and/or treatment of disease through individual protein sequences (personalized medicine).

Critical point theory is general, and recently it has proved to be the most effective way of describing protein networks that have evolved towards nearly perfect functionality in given environments (self-organized criticality). Evolutionary patterns are governed by common scaling principles, which can be quantified using scales that have been developed bioinformatically by studying thousands of PDB structures. The most effective scales involve either hydropathic globular sculpting interactions averaged over length scales centered on membrane dimensions, or exposed beta strand propensities associated with aggregative (strong) protein-protein interactions.

A central feature of scaling theory is the characteristic length scale associated with a given protein's functionality. To gain experience with such length scales one should analyze a variety of protein families, as each may have several different critical length scales. Evolution has functioned in such a way that the minimal critical length scale established so far is about nine amino acids, but in some cases it is much larger. Some ingenuity is needed to find this primary length scale, as shown by the examples discussed here. Often a survey of the evolution of the protein suggests a means of determining the critical length scale.

## 1. Introduction: Protein molecular complexity

Before we discuss the mathematical and statistical basis for protein scaling, we should first consider the complexity of the problem. A typical protein chain may contain 300 sites, each potentially occupied by one of 20 amino acids. The resulting number of possible amino acid sequences is $300^{20} \sim 10^{46}$, a number far larger than is normally encountered in physics, even in astrophysics. We shall later see biomedically important examples where a single site amino acid mutation is enough to change success into failure. Is it possible to explain such sensitivity using fundamental principles? Perhaps, although every protein family is different.

Several ambitious statistical studies of protein sequences are available. A broad evolutionary 2001 study of 29 proteomes for representatives from all three kingdoms: eukaryotes, prokaryotes, and archaebacterial, showed that proteins have evolved to be longer in eukaryotes, with more signaling ,heptad transmembrane helices in eukaryotes [1]. These are the signaling proteins which are the basis for about half of modern drugs. This very well-studied field is now classical, while research on most other proteins has just begun. A sophisticated group wish list (> 30 authors) of evolutionary problems associated with sequence evolution reported in 2012 that "current efforts in interdisciplinary protein modeling are in their infancy" [2].

Newtonian all-atom models are often unable to connect structural dynamics to function [3,4]. Evolutionary improvements are not easily recognized, as structures are seldom available for most species, and backbone structural differences are often too small to explain evolutionary progression. For example, the chicken and human peptide backbone coordinates of lysozyme *c* (Hen Egg White) are indistinguishable, even at highest attainable X-ray resolution in hundreds of PDB structures [5].

An historical note: science has grown rapidly in the last 60 years, and the science curriculum has struggled to keep pace. Here we often use the word "classical" to indicate topics that are old enough to have gained wide acceptance, and "modern" to describe the revolutionary changes that have taken place, often since ~ 2000. This review includes many modern ideas, which have enabled simple, powerful tools to organize molecular biology in new directions.

## 2. Mathematical tools

Mathematical and scientific literature, as measured by numbers of papers published each year, is growing super-linearly, with the number of specialized journals proliferating in the 21$^{st}$ century. Citations are widely used as a measure of the impact of research efforts, if not the absolute importance of their content. This growth has stimulated more than 700 papers analyzing the literature scientometrically. The most spectacular scientometric study so far surveyed nearly all the 20$^{th}$ century literature, consisting of 25 million papers and 600 million citations, with $10^6$ – $10^7$ more entries than typical data bases, and unique in the history of epistemology [6]. This study identified a citation transition, which occurred around 1960, and which is the earliest example of the cultural effects of globalization [7].

We can compare the development of different mathematical tools by using carefully chosen key words to search the Web of Science [8]. For example, although analysis is also a branch of mathematics, "analysis" is used in far too many non-mathematical contexts to be useful in a key word search. Useful mathematical words are algebra, geometry and topology; these yield surprising phase transitions over the last 30 years (Fig. 1). It is plausible that the glasnost emigration of Russian computer scientists around 1990 facilitated development of many biological data bases and contributed to the success of the human genome project. The fastest growing field, topology, is well suited to discussing protein network structure and evolution. One might have supposed that geometry, the oldest of the mathematical disciplines, would have stagnated by now, but modern geometry has evolved into differential geometry, a tool whose concepts are well suited to discussing the functions of sculpted globular proteins. The two-year 1989-1991 jump of special functions, a subclass of analysis, by a factor of > 20, is far larger than any of the general jumps in Fig. 1. Other subclasses also show large 1989-1991 jumps, for instance, "differential geometry", a factor of 9, and "network topology", a factor > 20.

## 3.   Statistical Mechanics

In his enormously popular Dublin 1943 lectures and book, "What Is Life?", Erwin Schrodinger proposed that we could progress in answering this question by using statistical mechanics and partition functions, but not quantum mechanics and his wave equation. He described an "aperiodic crystal" (today we would call it a glass) which could carry genetic information, a description credited by Francis Crick and James D. Watson with having inspired their discovery of the double helical structure of DNA [9,10]. Schrodinger arrived at this picture from thermodynamic theories concerning protein stability and information content. He observed that proteins are not only exponentially complex, but also must be near thermodynamic equilibrium, as they function nearly reversibly. Thus one can say that Schrodinger may have been the first theorist to conjecture that protein functionality could be usefully described thermodynamically.

Given what we know today, Schrodinger might have started more simply with the isothermal curves of the van der Waals equation of state (Fig.2). Apart from providing a quantitative description of the liquid-gas transition of many molecular liquids, this equation exhibits spinodal phase separation topped by a critical point on the critical temperature isotherm. The critical features of this model are common to many systems at or very near thermodynamic equilibrium. Specifically one can imagine that protein functionality can involve transitions between two end states, called "open" and "closed" by Karplus in his protein example [3]. Then if one knows from general considerations some quantitative properties of any protein sequence, one may be able to recognize extrema and critical points and quantify behavior in their neighborhood, without using elaborate Newtonian simulations. The comparative advantages of this more abstract method are its simplicity and universality, which makes it transferable. Such a method can easily treat evolutionary protein sequences by following their nearly critical properties, and may also facilitate identifying essential features in any protein and relating them to its functionality.

The importance of critical points in statistical mechanics has led to many studies of their properties in toy models, especially Ising models , which are lattice models with only nearest neighbor interactions. Ising himself showed his models for dimensionality d =1 do not contain a phase transition (a general property of all d = 1 models), and was unable to persuade himself that a transition existed for d = 2. L. Onsager studied critical point neighborhoods for a two-dimensional Ising model (1944), which he had solved exactly [11]. For d ≥ 4 the critical exponents of correlation properties are integers or half-integers (mean-field theory, used by L. Landau in his book on statistical mechanics). The most difficult case, d = 3, was solved using the renormalization group methods of particle theory. Its critical exponents are irrational sums of power series [12].

The existence of irrational exponents for d = 3 toy model phase transitions is suggestive, because empirical power law fits (which appear to be linear on log-log plots) are common in engineering discussions of nearly optimized systems. Power laws describe self-similarity (a power of a power is a power), and self-similarity is an appealing way of fitting together proteins over a wide range of length scales. Mandelbrot discussed geometrical self-similarity in terms of fractal exponents and power-law iterations (Mandelbrot sets) [13]. Per Bak connected these ideas to science in [14], and to evolution in [15].

## 4. Scaling

The general ideas discussed in 2012 in [2] and similar subsequent discussions of evolutionary dynamics and function [16,17] were unable to make contact with statistical mechanics and critical points. General ideas are all very well, but can they produce tangible biophysical and biomedically relevant results [18]? The positive answer to this challenge came from an unexpected direction, bioinformatic thermodynamic scales. The Kelvin temperature scale is central to entropy and general statistical mechanics, and T = 0 on the Kelvin scale corresponds to -273.16 C. Protein globular shapes are determined by competing hydrophobic forces (pushing segments inwards towards the globular cores) and hydrophilic forces (pushing segments outwards towards the globular water interface). Moreover, the leading physicochemical properties determining protein mutations are hydrophobicity, secondary structure propensity and charge [19].

To quantify these effects in the classic period of biophysics (before 2000), no less than 127 hydropathicity scales were proposed. Each scale had its merits, but few attempts were made to compare their accuracies, or applicability to properties other than those used in their definitions [20]. Meanwhile, the number and accuracy of PDB structures had grown enormously, creating the opportunity to re-examine the early geometrical definitions based on average neighboring hydropathicity volumes [21] or average surface areas [22] in Voronoi partitions [23] of proteins into amino-acid centered units with van der Waals radii. Note that the combination of Stokes' theorem and modern differential geometry suggests that there should be a close connection between the volume and surface definitions. The surface one in particular emphasizes hydrogen bonding to the water films which have shaped globular proteins in evolution, much as rocky coastlines are shaped by the pressures of tidal water waves [13,24].

With the stage set, in 2007 Brazilian bioinformaticists Moret and Zebende (MZ) built the interdisciplinary bridge connecting proteins to statistical mechanics and critical points [25]. They evaluated solvent-exposed surface areas (SASA) of amino acids in > 5000 high-resolution (< 2A) protein segments, and fixed their attention on the central amino acid in each segment. The lengths of their small segments L = 2N + 1 varied from 3 to 45, but the interesting range turned out to be $9 \leq L \leq 35$. Across this range they found linear behavior on a log-log plot for each of the 20 amino acids (aa):

$$\log \text{SASA}(L) \sim \text{const} - \Psi(aa) \, \text{const} . \log L \qquad ( 9 \leq L \leq 35) \qquad (3)$$

Here $\Psi(aa)$ is recognizable as a Mandelbrot fractal. It arises because the longer segments fold back on themselves, occluding the SASA of the central aa. The most surprising aspect of this folded occlusion is that it is nearly universal on average, and almost independent of the individual protein fold. Thus this striking universal result transcends and compresses thousands of individual protein folding simulations.

It is plausible that the MZ fractals exist because protein evolution has brought average protein SASA near critical points in each of the 20 aa subspaces. Remarkably, these subspaces span the same length frame $9 \leq L \leq 35$, independently of whether the aa is hydrophilic (near the globular surface) or hydrophobic (buried in the globular interior). One cannot "prove" with mathematical rigor these connections, but one can test (prove) them in the context of the evolution and functionality of many protein families. Comparing the results with those obtained with the classical scales enables us to estimate the relative merits of various scales.

When an {X(aa)} scale is available, it can be shifted linearly to a new {X′(aa)} scale

$$\{X'(aa)\} = a\{X(aa)\} + b \qquad (1)$$

without altering its relative values. Given the two constants a and b, one can arrange all scales to have the same average value and difference of largest and smallest values

$$<X'(aa)\}> = <\{X(aa)\}>, \quad \{X'(aal)\} - \{X'(aas)\} = \{X(aal)\} - \{X(aas)\} \qquad (2)$$

Moret and Zebende compared their fractal scale hierarchically to seven classical scales, the closest of course being [22], which utilized only the SASA of a single aa, averaged over a few entire protein structures available in 1985. Paradoxically, the MZ Mandelbrot fractals based on short segment log-log plots may be more significant, and not just because of critical points in configuration space. Both protein dynamics (on a scale of ms) and protein evolution (on scales as short as thousands of years) are difficult to understand in the context of their ~ $10^{46}$ configurational complexity (Levinthal's paradox) [26]. At first, it was not obvious why the MZ success occurred, but retrospectively we can see how fractal segmental character implicitly includes evolutionary optimization through exchange of modular building blocks [27]. Amino acid hydromodularity is apparently the best-documented example of effective parameter space compression [28].

Although [22] is closest to MZ, it is convenient to compare results obtained from the MZ scale (which implicitly describes second-order phase transitions), with those obtained from first-order protein unfolding measured by enthalpy changes from water to air (1982 KD $\Psi$ scale; this is also the most popular $\Psi$ scale) [29]. The differences between results obtained by the two hydropathicity scales should reflect improvements in accuracy, which could be biomedically important.

In a few cases hydropathic (inside/outside) $\Psi$ shaping may not be the most important factor, in which case we can turn to secondary structure (inside/outside) propensities of $\alpha$ helices and $\beta$ strands [30]. Hydrogen bonding is longitudinal for $\alpha$ helices, and transverse for $\beta$ strands, so we are not surprised to find that the (inside/outside) differences are small for $\alpha$ helices, and larger for $\beta$ strands which can bind outside. Protein binding can involve the $\beta$ strand exposed propensities for some cases. Another $\beta$ strand scale was constructed from the core sequence of amyloid $\beta$ [31], and we will return to this later. It is designed to emphasize aggregative "Hot Spot" propensities for binding on a length scale of 7 aa, and it will be referred to as $\beta$HS.

At first it may seem complex to have two hydropathicity scales $\Psi$ and two $\beta$ strand scales. Still, by comparing not only the scales with each other (through their Pearson correlation functions **r**), but also their performances on protein regions known to be structurally important, we will gain insights into molecular binding otherwise unobtainable. The best part is that comparative calculations with these scales are transparent and extremely easy, and can be implemented using only EXCEL spreadsheets, and their included software subprograms.

## 5. Critical Length Scales

In normal liquids critical opalescence couples long wave length light waves to long wave length density fluctuations. Near a critical point could long wave length ice-like film waves couple to long wave length solvent-exposed protein area fluctuations, thus explaining the origin of the MZ fractals? This is a difficult question. In the apparently simpler system of quenched (non-equilibrium) glasses there has been much theoretical discussion of the possibility of a diverging length scale at the glass transition [32], especially connected with long-range stress fields [33]. To discuss phase diagrams and physical properties of network glasses, one must start from specifics of their chemical bonds [34-36], as well as both the local and extensive topological properties, which include stress percolation.

Here we argue that biomedically important results are obtainable by judiciously combining specific length scales W = 2M + 1 with one or more of the Φ = {hydropathic Ψ/β strand scales}, denoted generically by Φ. Given Φ(aa), we calculate the modular average

$$\Phi(aa, W) = \text{average}(\Phi(aa - M), \Phi(aa + M)) \tag{4}$$

which is a rectangular window from aa -M to aa +M . It is possible to iterate this process, for instance one can easily see that

$$\Phi(aa, W, W) = \text{average}(\Phi(aa - M, W), \Phi(aa + M, W)) \tag{5}$$

gives a triangular window extending from aa -2M to aa +2M In practice for most cases rectangularly smoothed Φ(aa,W) appear to give best results. By using Φ scales we "dress" modular building blocks and enable protein interactions to appear to be short range with cutoffs.

Given that Φ(aa,W) is a good variable, how do we determine W? Experience with many examples suggests that all protein families are different, because their functions differ. We reflect some of these differences by choosing an optimized W, and others appear through comparisons between different components Φ. Because all four Φ scales have general meanings, comparison of Φ(aa,W) profiles often produces easily interpreted results. Far from being complex, the tools associated with Φ(aa,W) are powerful aides for exploring protein complexity. They automatically incorporate universal aspects of globular evolution.

## 6. Hinges and Pivots

Given a protein profile Φ(aa,W), one notices immediately that it has two kinds of extrema, hydrophobic maxima and hydrophilic minima. It is natural to suppose that the maxima act as pivots or pinning points for the conformational motion that is functionally significant, while the minima act as hinges. This language does not specify the conformational motion in Euclidean space that is functionally significant, as it jumps directly from the universal sequence geometries of Φ to function. What happens if we attempt to go elastometrically only between sequence and structure, by using the isotropic vibrational amplitudes of individual amino acids measured in structural studies?

The picture of differential aqueous sculpting of globular protein surfaces near a critical point [15] can be compared to elastometric treatments of hinge-bending conformational transition pathways [37-39]. A recent example is [40], which studied first-order open/closed conformational transitions. These are similar to the Ψ(KD) scale, which depends on first-order enthalpy changes between folded (in water) and open (in air). If there is a single mode that contributes significantly to the open/closed transition, it might have functional significance. However, the results showed only a wide range of overlap between conformational changes and single modes. It is unlikely that such changes or modes have functional significance. Although muscle contraction is mediated by a myosin cross-bridge which exists in two (open/closed) conformations, these were not explained by known conformers of myosin [41]. Instead there is an iterated hand-over hand motion of myosin along actin filaments [42].

## 7. Variance, Correlations and Level Sets

One of EXCEL's convenient software tools is variance, which can be used to quantify trends in extremal (phobic-philic) widths as functions of both the choice of Φ (for instance, ΨKD verses ΨMZ) and W. Informally variance is known as "mean of square minus square of mean", or

$$\mathrm{Var}(\Phi(aa,W)) = \Sigma\,((\Phi(aa,W) - <\Phi(aa,W)>)^2 = \Sigma\,\Phi(aa,W)^2 - n(<\Phi(aa,W)>)^2 \qquad (6)$$

where the sum is taken over n consecutive amino acid sites. In the context of the MZ Ψ scale, the variance measures the hydropathic roughness of the globular surface of the n sites of the protein chain segment. The local or global roughness can affect dynamical functions, which should occur neither too fast nor too slowly, in order to synchronize with other protein motions [43]. Mixtures of rougher granules have lower packing densities, and the granular knobs can jam kinetics [44].

Variance is a useful quantity in studying protein evolution and dynamics because it combines extremal hydropathic pivots and hydrophilic hinges on an equal footing. One might suppose that such a simple function would have been used for long times in biology. In fact its bioinformatic importance was first realized only in 1911 by R. Fisher (then a student), but its publication centenary still lies ahead, in 2018. Fisher used it to describe "The Correlation between Relatives on the Supposition of Mendelian Inheritance". It conveniently represents the random combinations of parent genes.

Pearson correlations themselves are normalized cross variances between two functions, for instance two sequences of the same protein from different species or strains X and Y ($-1 \leq r \leq 1$)

$$r = <(X(aa)\} - <X(aa)> \ (Y(aa) - <Y(aa)>)/(<(X(aa))^2> - <X(aa)^2>^{1/2} \ >) \ (<(Y(aa))^2> - <Y(aa)^2>^{1/2} \quad (7)$$

Level sets were developed to track the motions of continuum interfaces [45] – applied here to protein globular surfaces. Mathematically oriented readers will find "simple" explanations of their background and comparative computer science advantages online, for instance, under "Level Set Methods: An initial value formulation". Practical applications of level sets have emphasized image analysis [46,47], and have gradually evolved to include Voronoi partitioning, just as has been used for deriving protein hydropathicity scales since 1978 [48,49]. We expect, of course, that hydrophobic pivots move most slowly, while hydrophilic hinges move fastest. When there are two or more level pivots or hinges, it is likely that this is not accidental (nothing in proteins is), and we can test this assumption by comparing profiles with different scales, the KD $\Psi$ and MZ $\Psi$ scales, for example.

## 8. Evolution and Mutated Aggregation: Hen Egg White

Lysozyme *c* (aka Hen Egg White, or HEW) was for some time the most studied protein: the PDB contains more than 200 human and 400 chicken HEW structures. HEW is also present in many other species, not only in the 400 million year old chicken sequence, but in most other vertebrates, almost unchanged in its peptide backbone structure. The backbone structure is exceptionally stable, with human and chicken $C_\alpha$ positions superposable to 1.5A°, while the aa sequence mutates from chicken to human with 60% aa conservation [10], well above the 40% minimum usually necessary for fold conservation [50].

HEW is a comparatively small 148 aa protein, which has a nearly centrosymmetric tripartite α helices (1- 56 and 104-148) and β strands (57-103) secondary structure. During its long career, HEW has performed at least three functions, as an enzyme, an antibiotic, and an amyloidosis suppressor. The relative importance of these functions has changed from species to species, and it seems likely that these changes are reflected in the amino acid mutations that have maintained the centrosymmetric structure.

Because we are most interested in the evolution of the properties of HEW from chickens to humans, we plot the roughness or variance $\Re$ as $\Re_S(W)/\Re_{Human}(W)$ in Fig. 3, using the fractal MZ $\Psi$ scale, for a range of species S. There is a broad peak, together with a narrow peak, both centered on W = 69, roughly half the protein length. The peak also occurs at W = 69 with the KD $\Psi$ scale, but its amplitude is only ~ 60 % as large, so the roughness evolution is better described as thermodynamically second order. Similar narrow-broad peak structures have been observed in critical opalescence spectra, where the broad peak, associated with phonons, is called the Mountain peak [51]. Note that this "universal" peak applies to the terrestrial species, but not to zebrafish.

The structure-function relations giving rise to these peaks can now be profiled with W = 69, as shown in Fig. 4 for the extreme terrestrial cases of chicken and human. Note that normalizing $\Re_S(W)$ by $\Re_{Human}(W)$ in Fig. 3 is natural, because human structures have evolved to be closest to critical and smoother. This means that in most cases the critical limiting behavior of ideal functionality is nearly reached with the human sequence, described partly by $\Re_{Human}(W)$. It is obvious from Fig. 4 that on the W = 69 length scale, which is approximately half the HEW length, the largest effect of evolution has been to stiffen the flexible central β strands by making them less hydrophilic.

While the results shown in Fig. 4 dramatically confirm the previously hidden content of HEW amino acid sequences, what is it? The characteristic length scale for amyloidosis is W = 40, because this is the length of the Aβ fragment (also discussed in Section 13) responsible for forming amyloid fibrils. On this length scale one can examine the effects of mutations on HEW aggregation rates, and a detailed discussion shows consistent shifts [50]. Presumably amyloid suppression is a key function for advanced species with larger neuronal networks which must be stable for longer lifetimes.

One can also discuss mutated aggregation rates in a similarly centrosymmetric protein, measured in [19], and analyzed there using W = 5 only (no other values of W were considered). The 98 aa α/β protein acylphosphatase (AcP) resembles HEW (αβα) in that its sequence is still nearly Ψ centrosymmetric, but it is more complex with five regions instead of three (β<1-19>, α<20-32>, β<33-54>, α<55-70>, β<71-98>). A parameterized W= 5 method for studying HEW mutations was applied to AcP, with disappointing results: increases in mutated aggregation rates were expected only in the central region, and found only in the N- and C- terminal wings [19].

When the S/human roughness ratios are plotted for Acyl-1,modular the results are much more complex than in Fig.3 for HEW. The difference is attributed to its richer α/β structure [50]. The overall scale, as measured by chick/human, is about 30 % enhancement, which is about 10 times smaller than that shown for HEW in Fig. 3. Nevertheless, three features were easily identified. Chick/Human gave peaks at W = 43 and W = 25, as well as a human peak at W = 13. These three values of W probably reflect interactions with three other (unknown) proteins. One can assume that mutations of the human sequence tend to "undo" evolutionary improvements and cause mutated human profiles to regress towards chicken profiles. The three profile difference patches then agree well with the patterns of mutated aggregation rates [50].

The aggregation of globular proteins, such as well-studied lysozyme *c* (Hen Egg White), may involve unfolding, and is thus more complex than that of Aβ, a known product of A4 fragmentation. The smallest lysozyme amyloid nucleus is 55-63 (9 aa) GIFQINSRY, called K peptide [52,53]. K peptide is the strongest amyloid former of nine related small (< 9 aa) peptides over a pH range from 2 to 9. Profiles for entire human lysozyme in Fig. 4 show that the 9 aa K peptide nucleus is located at the center of centrosymmetric α-β-α lysozyme. The 69 aa wide central β region 57-103 is hydropathically level, so its β strands are nonamphiphilic [50].

## 9. Level Set Synchronization: the Case of HPV Vaccine

Synchronized motions of actin cell skeleton proteins guide cell surface and interface deformations, a protein realization of Stokes' theorem that also involves criticality [54,55]. Most proteins have > 300 aa, and their functionality at the molecular level also involves large-scale conformational motions which are optimized by synchronization. The first example of molecular level set criticality is likely to come as a surprise, as it concerns a single amino acid mutation of a 505 amino acid protein, which alters the self-assembly rate protein complexes of HPV cervical cancer vaccine by a factor of $10^3$ [56].

The long road that led to cervical cancer vaccines began in 1976 when Harald zur Hausen published the Nobel hypothesis that human papilloma virus (HPV) plays an important role in the cause of cervical cancer. HPV is a large capsid protein, but it was found that only the 505 aa L1 part was needed to make a good vaccine that conformationally self-assembled into morphologically correct virus-like particles (VLPs). L1 from HPV 16, taken from lesions that had not progressed to cancer, self-assembled $10^3$ times faster than the HPV 16 L1P that researchers everywhere had been using; the old strain L1P had been isolated from a cancer, which differed from L1 by only a single amino acid mutation D202H [56]. The huge increase in self-assembly rate could well be due to conformational synchronization, but this is not easily quantified using Newtonian methods.

Given the lower bound of L = 9 in (3), and the remarkable properties of L1, its profile Ψ (aa,9) with the fractal scale (3) was examined near the 202 mutation site [54]. The striking feature is the presence of two almost level L1 hydrophobic peaks in the region between 191 and 231, shown enlarged in Fig. 5. The narrow peak α is centered near 202, the mutated site distinguishing L1 from L1P. The level condition is satisfied to within1% by L1, but by only 5% by L1P and by two other singly mutated strains recently added to the PDB. Note that no mutations were found in the stabilizing broad peak. Note also the deep hydrophilic minimum near 215, which functions as a plastic hinge accelerating self-assembly.

This example also brings out the advantages of scaling with Φ(aa,W). The excellent agreement shown in Fig. 5 disappears when W is reduced to 7, below the cutoff in (3). It also disappears when ΨMZ is replaced by ΨKD. In other words, the pre-2000 efforts involved in constructing 127 different Ψ scales were involved in a good direction, but proteins are so complex that success was possible only bioinformatically after the PDB structures became numerous and more accurate [57].

In addition to the types of HPV that cause cancer, there are "milder" types that cause only warts (self-limited growth). It might appear that the differences between these two types, which occur on a cellular level, could not be analyzed on a molecular level. However, there are many self-similar aspects to proteins and cells, so one should look at the differences between the HPV16 (cancer) and HPV 6 (warts) profiles (Fig. 6). There are large differences in the amide (N) – terminal region, far from the plasticity hinge (which is almost unchanged) seen in Fig. 5. Thus the main self-assembly function is unchanged, but the N-terminal region changes can account for the sometimes serious side effects. These small differences are much reduced with the classic KD scale, which is suited to some large open-closed transitions, but not small conformational changes [57].

## 10. Level Set Synchronization: the Case of the False Aspirin

Hippocrates (~ 400 BC), the father of modern medicine, mentioned the miracle drug aspirin as a powder made from the bark and leaves of the willow tree [58]. By 2000 chemists knew the molecular mechanisms of aspirin (acetylsalicylic acid) [59,60]. Aspirin and other non-steroid anti-inflammatory drugs (NSAIDs) inhibit the activity of cyclooxygenase (COX-1) which leads to the formation of prostaglandins (PGs) that cause inflammation, swelling, pain and fever. However, by inhibiting this key enzyme in PG synthesis, the aspirin-like drugs also prevent the production of physiologically important PGs which protect the stomach mucosa from damage by hydrochloric acid, maintain kidney function and aggregate platelets when required. This conclusion provided a unifying explanation for the therapeutic actions and shared side effects of the aspirin-like drugs, which when overused, cause ulcers.

A unique example of a natural enzyme twin was discovered in the early 1990's, COX-2. It has nearly the same length (~ 600 amino acids) as COX-1, but is encoded by a different gene! BLAST sequence comparison of COX-1 and COX-2 showed 60-65% sequence identity of COX-1 and COX-2, which should be enough to prevent elucidation of their functional differences from structural data alone [60], even when augmented by Newtonian simulations (which have so far not been attempted).

COX-2 inhibitors became attractive when it was realized that they could have the same anti-inflammatory, anti-pyretic, and analgesic activities as nonselective inhibitors NSAIDs, with little or none of the gastrointestinal side-effects [60]. An inhibitory COX-2 drug passed through all its tests, and was duly marketed for several years. Then it was discovered that COX-2 inhibitors caused the appearance of significant cardiovascular toxicity associated with chronic use (2–4% of patients after 3 years).

What causes the difference between COX-1q and COX-2 inhibitors? This difference remains unexplained, although there are few enzymes of lipid biochemistry for which there is such a wealth of structural and functional information [61]. However, this structural information is limited to COX complexed with inhibitors, which are known to strain the wild structures [62]. In other words, here (as often) the ~ 600 amino acid structural differences are both small and complex. Enzyme inhibition occurs in transition states (saddle points in an energy landscape), which are conformationally different from the ground state minima of a bound complex. Moreover, as we shall see, there are indications that many large protein interactions occur near cell membranes, which orient the conformational distortions in the transition states.

Because of the large sequence identity of COX-1 and COX-2, it may appear difficult to analyze their differences quantitatively. However, their similarity can be turned to advantage by comparing their overall roughnesses or sequence variances (see (6)). The $\Re(W)$ differences are small for small W, but increase with increasing W, finally peaking at W = 79 [64]. We can interpret this large value of W as an average spacing between the centers of modules that are involved in critical large-scale conformational changes that distinguish COX-1 from COX-2. A close inspection of the sequences aligned with BLAST suggests splitting each type into three domains, A: 1-110, B:110-290, and C: 290-end, but this separation is more obvious from the hydropathic profiles. [64] repeated these calculations with the KD scale.

The general result from comparing the MZ9 and KD9 profiles is that the MZ9 scale is much more accurate, and the central domain B:110-290 is most scale-sensitive. There are many quantitative differences between the MZ and KD profiles [64], which could have a direct bearing on self-assembly kinetics. The greater precision of the MZ scale is expected, because all the differences must be conformational (large-scale and small). The differences could be explored with mutagenesis studies of self-assembly kinetics in the type-sensitive regions.

## 11. Globins and the Correlated Evolution of Metabolism

Myoglobin (Mb) and neuroglobin (Ngb) are two globin-based proteins responsible for oxide-binding in tissue. As their names would suggest, Mb is typically found in muscle tissue where it stores oxygen [65], while Ngb is concentrated in the nervous system. While globins have typically been thought to have respiratory purposes, recent studies of Ngb have revealed that they also can have a number of other physiological functions [66]. These physiological functions are closely related to the structures of the oxygen channels of the proteins. Further analysis of how the oxygen channels differ across species of Ngb and how they differ from the oxygen channel in Mb helps understanding these physiological functions.

At the center of each globin is the heme unit, which consists of Fe bound to the center of porphyrin, a large planar ring composed of 4 smaller C- linked rings.  The Mb structure (~ 150 amino acids) was the first protein structure to be determined (in the 1950's), and discussion of its structure-function relations continued well into the 21$^{st}$ century.  Perhaps the most interesting question is the nature of the oxygen channels which control $O_2$ and CO passage.  These channels contain a sequence of pockets, observed in Xe-doped samples.  Herculean Newtonian simulations have shown that the $O_2$ and CO ligands have a choice of two channels, from opposite sides of the porphyrin [67,68].  The protein chain wraps the porphyrin macro-ring, and passage through the two closest pockets is controlled by Histidine (pKa = 7, neutral) gates.  The spacing of these gates is 29 amino acids in Mb, while in Ngb, the spacing is 32 amino acids.  This spacing is nearly constant across all terrestrial species, although its amino acid composition varies.

Given the key role played by the 29 amino acid His gate spacing, it is natural to profile both human Mb and Ngb using W = 29 [69], as shown in Fig. 7.  There are some similarities between the two profiles, but there are also striking differences, which are directly connected to differences in function (see Fig. caption).  A detailed discussion [69] shows that the principal Euclidean features of the different oxygen channels of Mb and Ngb correspond quite accurately to the fine structure of these profiles.

A striking feature of Fig. 7 is the approximate three-fold symmetry of the human Ngb profile, with three nearly level and equally wide hydrophobic pivots, and two nearly level hydrophilic hinges.  This is somewhat similar to lysozyme (Fig. 4), but here the symmetry is even more pronounced.  Is this accidental?  Fortunately we have Ngb sequences for many species, enabling us to study its evolution from cold-blooded aquatic species across the evolutionary panorama up to humans.   The simplest case is tropical and temperate freshwater fish, shown in Fig. 8.  In tropical zebrafish, the Ngb profile is very nearly tripartite level, while in temperate fish the structure is stabilized by hydrophobic wing peaks.

Because Ngb performs the most demanding metabolic functions − supplying oxygen for neural and retinal signaling − its remarkably balanced profiles are to be expected from its evolution. Mb shows one spectacular feature in its evolution from chickens to humans. Mice are favorite meals for predators, yet they flourish because of their ability to escape rapidly (a favorite theme for cartoonists!). To do so, their muscles must be supplied with oxygen rapidly. Comparing mice with chickens, we see (Fig. 9) a striking difference, associated with the appearance in mice of a hydroneutral peak associated with the apex (center) of the 29 amino acid chain connecting the two His gates.

The critical role played here by the His gates is an intrinsic part of the fractal MZ scale. The stability of histidine makes it the central and most conserved element of many catalytic triads [70], the most studied examples being Serine-Histidine-Aspartate (chymotrypsin) and Cysteine-Histidine-Aspartate. Catalytic triads form a charge-relay network (central His has pKa = 7), and are excellent examples of convergent evolution [71].

The standard scale for mutation rates (BLOSUM 62, used in BLAST) exhibits a deep hydroneutral minimum in mutation rates near its center [72]. With the MZ scale this minimum is associated with alanine (A), glycine (G), the smallest amino acid, and histidine (H). Table I of [25] shows that none of the older scales places all three of these amino acids at its center. In terms of RMS deviations from the average value of each scale, the off-center differences are 7 times or more larger for the other scales than for the MZ scale. Note that these differences are not much larger than the quoted error bars in the classical work; the modern fractal MZ scale, based on an L = 20 centered range of solvent accessible areas, is more accurate in principle, but it also benefits from its post-2000 bioinformatic survey of more than 5000 segmental structures.

Hemoglobin is the classic protein oligomer. It colors red blood cells, and it transports oxygen from the lungs throughout the body. It is a dimer of dimers based on Mb-like monomers. Whereas Mb and Ngb ligand interactions involve charge, the Hb tetramers absorb and release oxygen primarily through interactions between strain fields localized near hemes and extended strain fields associated with dimer interfacial misfit [73]. Strain fields are the microscopic mechanism for long-range ("allosteric") protein interactions, because charge interactions are screened by the large dielectric constant (~ 100) of water. Allosteric aspects of globin functionality, especially the Christian Bohr (father of Niels Bohr) cooperative oxidation of tetrameric Hgb, have been the subject of 6500 papers [74].

Simply by comparing the profiles of the Hb dimers ($\alpha,\beta$) with the Mb and Ngb profiles, we see immediately that Hb strain field interactions are synchronized by level extrema, much like the Ngb interactions, and qualitatively different from the Mb interactions, which are dominated by a single hydrophilic hinge (Figs. 7,9,10). Strain fields - their observation and theoretical quantification – are one of the most difficult areas of condensed matter physics (for example, microscopic theories of cuprate high temperature superconductors [75]). When two fields interact, their spectra often exhibit (in the atomic case, Fano) antiresonant interference dips following resonant peaks [76,77]. An antiresonance is observed in Hb ($\alpha,\beta$) correlation (see equation (7)) spectra [73]. The importance of strain fields in the globins suggests amusing similarities between globin functionality and superconductivity (strain-field generated Cooper pairs).

## 12. Self-Organized Criticality and Color Vision

Next to the neural network itself, three-color vision (with depth perception!) is perhaps the most complex living function. Rhodopsin is the primary color opsin, and comparison of various species shows human rhodopsin has the smallest hydropathic MZ variance, that is, it is the smoothest. This suggests that its function is probably best preserved for long times by recoil involving small rotations with little wear and tear. The relative smoothness of human rhodopsin increases with increasing W [78]. One might suppose that other color opsins are also smoothest in humans, and this is true for most species. However, cats are nocturnal predators, who rely on their red opsins, and their red opsins are the smoothest. Dogs, who hunt in packs, have inferior vision [79].

All retinal frameworks have a rod-and-cone structure. There is a broad evolutionary difference between vertebrate and invertebrate rhodopsin roughness profiles $\Re$(W). The small-scale cellular structure of compound invertebrate eyes is less susceptible to cumulative amplification of lateral instabilities (inter-rod slippage), which are suppressed in large single-cell vertebrate retinas by rougher rhodopsin. This enables insects to utilize smoother rhodopsin, which probably enhances their visual temporal resolution (200 images/sec in bees, compared to 30 images/sec in humans), as their optical responses are less slowed by multiplied inter-rod constraints. Invertebrates have compound eyes with 2 primary cells and ~ 10 secondary pigment cells per retina, compared to ~ $10^5$ rod cells/retina in mammals [80]. This difference corresponds to a change in lateral molecular correlation lengths (average retinal size) of a factor of 100.

The smoother invertebrate roughness profiles $\Re(W)$ for $W < 25$ reflect a trade-off between long-range smoothness for $W > 25$, which is ineffective with only 10 pigment cells/retina, in favor of smoother photoreceptor-membrane interactions in their ommatidia. It is striking that the retinal cellular morphological differences (single/compound lens) between vertebrates and invertebrates are qualitatively obvious in their rhodopsin molecular roughness profiles $\Re(W)$. In principle the dominance of $\Re(W)$ for $W > 25$ in large mammalian retina reflects the nearly scale-free property at large lengths that characterizes self-organized criticality in more recently evolved species.

Empirically it is known that color discrimination (or quantization) can be described by three 8, 8 and 7 bytes/color channel, and that this discrimination is almost perfectly multiplicative, leading to a rough estimation of the number of distinguishable colors in the optimal color space as $3.2.10^6$ (22 bytes) [81]. Retinal organization in primates, which have a complex visual behavioral repertoire, appears relatively simple. However, quantitative connections between primate vision, retinal electronic connectivity (or neural network) and color quantization have so far surprisingly not emerged from studies of physiological models [82].

Balance between competing channels is characteristic of many dynamical critically self-organized networks [83], and is independent of chemical details. The networks are nearly optimized near self-organized criticality, and their dynamics follows another extremal principle, maximum rate of entropy production [84], in this case, fastest download and reset utilize the same number of degrees of freedom as activation. The separation of visual signals into three channels will be most efficient if those channels are nearly equally informative, and this is found to be the case [85]. This implies that the primate 20 aa opsin sequence is also nearly optimally designed to interact with all three pigment download paths, and that these paths can be primarily influenced by opsin aa hydrophobicities.

To count the number of independent optoelectronic channels accurately, one should make allowance for amino acid hydroredundancy. Even if the MZ hydrophobicities are not well separated, two aa may still make distinctive contributions if their strongly $\pi$ polarizable ring contents are different. Combining these two criteria, hydroredundancy occurs in only three cases: for one-ring Tyrosine ($\psi(Y) = 0.222$) and Phenylalanine ($\psi(F) = 0.218$); aliphatic chain (no rings) Isoleucine ($\psi(I) = 0.222$) and Methonine ($\psi(M) = 0.221$), and Alanine ($\psi(A) = 0.157$) and Glycine ($\psi(G) = 0.156$). Thus there are effectively N = 17 remaining functionally independent aa available, with a byte content of 4.12. The number of possible combinations N! of these channels is given by Stirling's formula (logN! = NlogN – N).

The visual system is optimized to monitor moving objects, and this action involves resetting the retina to its dark state after each observation, in order to be ready for the next observation. This two-step ON/OFF process with nearly optimal two-step serial uploading and downloading must resolve 2(4.12) = 8.24 ~ 8 bytes/tricolor channel, which is in very good agreement with the empirical CIELAB color difference formula (8,8,7) [81]. Note that this two-step reversing model is easily countable, while two-step chemical mechanisms (phosphorylation, etc.) are not. The two co-exist, but in terms of shaped globular amino acid configuration space, the mechanical model is more fundamental [86]. The interface of protein structural biology, protein biophysics, molecular evolution, and molecular population genetics forms the foundations for a mechanistic understanding of many aspects of protein dynamics [2].

### 13. Platelet Aggregation Inhibitors

Hemodialysis is used to treat patients with total or partial kidney failure, which results in protein aggregation. Several blood-based biomarkers have been tested to monitor partial kidney failure [87, 88], a globulin and an enzyme. We can compare their profile evolutions [89]. β-2 microglobulin (β2m) is a small (119 aa) single-domain protein, characterized by a seven-stranded β-sandwich fold typical of the immunoglobulin domain family. We see in Fig. 11 how β2m has evolved from chicken to human, with large amino acid changes in the partially conserved (17-116) region, Iden., 48% , Posit 67% (BLAST), with enough similarity that the fold is still conserved. The feature that has evolved most is the leveling of the two central hydrophobic peaks. This small protein assists in removing molecular refuse from kidneys [20,21], and it can best do this by synchronizing its C- and N- halves. This is more important in long-lived humans, so on going from chickens to humans, β2m has evolved level hydrophobic peaks. Similar leveling of hydrophobic peaks differentiates the evolution of myoglobin (154 aa, larger because it is wrapped around the heme).  Compare especially the leveling of Mgb hydroprofiles of tropical and temperate freshwater fish, Fig. 8.

The 190 aa enzyme prostaglandin-H2 D-isomerase (P2GDS) has evolved much more rapidly than β2m, and the human and chicken folds are different [63]. The results of recent evolution from rat to human (conserved fold, Iden., 69% , Posit 80%) are shown in Fig. 12.  Both the hydrophobic peaks and the hydrophilic dips are much more level in the human profile.

## 14. Evolution of Flu Glycoproteins

Influenza is an RNA virus with a high mutation rate, nearly one mutation per replication [90]. It is also highly infectious, extending across an effective population size of donor-recipient pairs estimated to be approximately 100-200 contributing members [91]. This high mutation/transmission rate facilitates rapid selection for viruses that succeed in evading both antibody-mediated immunity and vaccines, and may involve ~3-5% substitutions in the antibody-recognized regions.  Traditionally viral evolution (sometimes called antigenic drift and shift) has been represented by phylogenic trees [92], but the utility of this representation has been questioned on fundamental grounds [93,94].  Trees are dimensionally limited, as their effective dimension d is d = 1+ δ, where δ << 1 is the average rate of formation of new branches. Epidemics occur when a new evasive strain cluster emerges in a minimal mutation space with dimension d ≥ 2 [93].

The public data base for flu strains amino acid sequences contains hundreds of thousands of strains, is probably the most extensive in protein science, and so should be highly suited to analysis by scaling theory.  The ultimate goal is an effective and timely determination of suitable target strains for vaccine development.  Because hundreds of millions of flu vaccines are produced annually, this goal challenges most methods based on general considerations only [93].

There are several types of flu viruses, and the current vaccine cocktail treats the most common three.  The oldest type A/H1N1 was responsible for both the deadly 1918 pandemic and the unexpectedly harmless 2009 "swine flu" pandemic. The H1N1 type may be as much as 500 years old, and it appears to have stabilized, with δ << 1.  The success of traditional methods based on phylogenic trees for H1N1 left virologists unprepared for vaccine failure (effectiveness reduced from 50% to 15%, with higher morbidity and mortality) associated with the 2013-4 A/H3N2 strains, which are "young" (only 50 years old) [95].

Influenza virus contains two highly variable envelope glycoproteins, hemagglutinin (HA) and neuraminidase (NA). The structure and properties of HA, which is responsible for binding the virus to the cell that is being infected, change significantly when the virus is transmitted from avian or swine species to humans. HA is used to make vaccines, and its variations can be measured antigenically (through time-consuming ferret blood responses). Scaling theory quantifies the simpler problem of the much smaller human individual evolutionary amino acid mutational changes in NA, which cleaves sialic acid groups and is required for influenza virus replication. These two glycoproteins combine to form spikes on the nearly spherical viral surface.

When one studies the panoramic evolution of NA variances, a striking trend emerges (Fig. 13). The strains have not evolved randomly, as is often assumed. Instead, the N1 strains became overall smoother as they evolved, except for occasional outbreaks [96]. This smoothing enabled the strains to evade antibodies generated by vaccines, but they also made the virus less virulent. The long-term benefits of sustained vaccination programs shown in Fig. 13 are large and are not easily recognized in short-term studies [97]. Meanwhile, what was happening to H1?

The HA1,3 story has turned out be too complex for treatment by scaling theory [98], and is discussed in the following paragraphs. The NA2 story has turned out to be quite interesting, as it explains why H3N2 mutates so rapidly that it escapes phylogenic predictions, and why it is more virulent than H1N1. We can compare the hydroprofiles of NA1 [96] and NA2 [95]. There is a deep hydrophilic minimum of N2 near 335, which is associated with a wide disordered 37 amino acid region (no α helices or β strands). The three deepest hydrophilic minima of N1 are less deep, and are associated with shorter disorder: (150), 17; (220), 15; and (390) 13 − less than half as large disordered regions (PDB 4B7N). The deep hydrophilic minimum of N2 creates a large, exposed and elastically soft surface region, which is both easily mutated and can flexibly enhance cell penetration. The interactions of the two glycoproteins HA and NA are strongly coupled, and it is N2 that "drives" the dangerous H3N2 strains.

Although it is outside the scope of this review, Deem et al, using heuristic methods [99,100] (inspired by modern statistical mechanics [101]) on the large sequential data base appear to have made the greatest progress, starting with a seminal H3N2 paper in 2006, when greatest interest was focused on H1N1. They found that simply compressing the H3 sequence mutations into a 2-dimensional principal component space was enough to yield a reliable picture that identifies new strain clusters, without involving the slow and low-resolution ferret antigenicity measurements that are used to monitor vaccine safety. They also showed that vaccine effectiveness is most easily predicted from mutations of nearly conserved sites on the HA heads, which form distinct subsets called epitopes. Here the binding function of HA switches from one epitope to another, and the most effective vaccines follow switches. Fig. 14 is a recent d = 2 map showing punctuated creation of H3N2 strain clusters in timely 6 week periods [102].

## 15. Secondary Thermodynamic Scales

Protein globular shapes are determined by competing hydrophobic forces (pushing segments inwards towards the globular cores) and hydrophilic forces (pushing segments outwards towards the globular water interface). Differential (fold-conserving) changes in globular structures which involve hydropathic extrema are essentially topological, and correspond to inside/outside structural features. Differential functional properties are usually determined by outside (surface) mutations, as the inside amino acids are more densely packed, and not easily rearranged while conserving the fold.

The leading physicochemical properties determining protein mutations are hydrophobicity, secondary structure propensity and charge [19]. Secondary structure consists of $\alpha$ helices (like Myoblobin) and $\beta$ strands (amyloids), which for our purposes can be regarded as longitudinal and transverse H-bond networks. The inside/outside bifurcation has been developed for of $\alpha$ helix and $\beta$ strands in two ways: bioinformatically by surveying > 2000 structures, labelled FTI [103], and by a Hot Spot model based on central mutations of a central $\beta$ amyloid 7-mer segment (labelled HS) [104]. FTI found, as one would expect, that the inside/outside propensities were small for $\alpha$ helices and large for $\beta$ strands.

Comparison of these inside/outside scales shows Pearson correlations r ~ 0.9. This means that small differences between the scales can have large functional consequences. Some examples are discussed in [105]. A common example of an all-β protein is Tenascin, an extracellular matrix glycoprotein containing 14 repeats and > 2000 amino acids. It has many functions, including guidance of migrating neurons as well as axons during development of neural networks.

## 16. Holy Grail of Cancer

Vincent DeVita was instrumental in developing combination chemotherapy programs that ultimately led to an effective regimen of curative chemotherapy for Hodgkin's disease and diffuse large cell lymphomas. Along with colleagues at the NCI, he developed the four-drug combination, known by the acronym MOPP, which increased the cure rate for patients with advanced Hodgkin's disease from nearly zero to over 70%. This was the first cure for a solid tumor. The MOPP drugs are simple and inexpensive, and a dramatic demonstration of the power of combinational chemotherapy [106].

The MOPP drugs separately are less effective. Synergistic cancer drugs are often described in terms of checkpoints in the native immune system. There are >20 different ligand–receptor molecular checkpoint interactions between T cells and antigen-presenting cells [107]. Drug treatments of most of these reactions separately are ineffective, and often depend on the individual set of cancer mutations. There are > 10 million amino acid mutations associated with cancer, so extending combinational chemotherapy to the molecular level is a challenging problem, not so much for computers, as for building an adequately large DNA data base to be treated computationally.

If a simple, cost-effective cancer biomarker for early cancer detection (ECD) could be developed, it would immediately enable many cancers to be treated effectively using known and also cost-effective methods. It would also make building a combinational chemotherapy checkpoint data base much easier, so DeVita has called such a biomarker the Holy Grail of Cancer [106]. At present the leading candidate for such a marker appears to be the 15-mer epitopes selected from 400 aa p53 and mucin in the only large-scale (50,000 patients, hence cost-effective) clinical studies reported so far [108]. The discovered epitopes are more sensitive for ECD than p53 itself, which is more sensitive than any other protein.

Small antigenic regions (epitopes, as small as 10 aa) interact with small antibody regions (paratopes). Small peptide epitope sequences are printed cost-effectively on microarrays. Due to their miniature format they allow for the multiplex analysis of several thousands of peptides at the same time while requiring a minimal sample volume [109]. The most famous and best-studied (~ 100,000 articles) protein is the "cancer suppressor" p53 (~ 400 aa). 15-mer p53 epitope scans [108] revealed a bifurcation in the autoantibody pool between early cancer and tumor populations, which is the critical test for ECD. The tumor epitope lies near the center of p53, while the most sensitive early cancer epitopes lie in the N- and C- terminal wings. This central/wing pattern occurred in the mutations of HEW and AcP (Section 8), and is predicted correctly only by using the best MZ hydropathicity scale.

Can we use scaling to confirm the (early cancer)/tumor||wing/central parallel dichotomy epitope sensitivity structure for p53? The answer is both No and Yes. Hydropathic $\Psi$ scaling does not identify any of the epitopes discovered in [108]'s epitopic scan, but scaling with secondary (Sec. 15) exposed β strand propensities [103] is successful [110]. Most proteins are hydropathically compacted into globules, but the tumor suppressor p53 (393 aa) forms a flexible, tetrameric four-armed starfish [111], quite distinct from the globular structures which most proteins (even when oligomerized) form. Among all proteins p53 is much more hydrophilic than average, and it is also elastically much softer, with about half its structure dominated by β strands, while the remainder (especially the N-terminal quarter) is disordered.

The most interesting ECD example of the fit to the experimental 15-mer overlapping epitopes numbered XY, containing amino acids $5(10X + Y) - 4$ to $5(10X + Y) + 10$, is XY = 09,10. The profile in Fig. 15 shows that overlapping 15-mer epitopes -9 and -10 were successful because they share a common 7-mer. The profile -9,-10 obtained with the FTI βexp scale is much more useful than that with $\psi$ hydropathicity scales [110]. Such comparisons are themselves simple and easily applied "checkpoints" for testing the theory.

A subtle point in the scanning profiles is the weakness of the XY = 34 central epitope sensitivity in early cancer stages, even though it is the only dominant epitope in the presence of tumors [108]. Initially the immunogenic interactions may be weaker and longer ranged, so that -34 does not initially form β strand bonds to antibody paratopes, whereas -9,-10 does. At later cancer stages, the interactions could be stronger and shorter ranged, with entangled interactions involving nonlinear amino acid interactions. Such nonlinear interactions are not easily connected to DNA personalized mutations, whereas such connections may well be possible with the weaker interactions discovered in ECD. Not only are drugs addressing weaker interactions much more likely to be effective, but they could also be much more easily identified.

The superior biomarker sensitivity of p53 epitopes to p53 itself is understandable, considering that the autoantibody paratopes probably form families themselves (so far unknown). The sensitivity of the individual p53 epitopes is ~ 30%, but with better statistics and correlations with dynamically monitored individual p53 mutations it could be higher. In any case, mucin is another whole protein cancer biomarker that can be used as a source of autoantibody detective epitopes, which can probe different autoantibody families [112,113]. The structure of mucin (~1250 aa) is dominated by 20 aa repeats, whose length increases from 16 in mouse to 30 -90 in humans. The ~ 1000 aa repeat domains are more stable in the center and more variable in the periphery, so [112,113] found that the most sensitive mucin epitopes consisted of three center repeats; presumably the two outside repeats buttress the autoantibody binding center repeat.

Although the mucin structure is disordered, it can still be profitably analyzed with scaling theory, using both hydropathic $\Psi$ and transverse $\beta$ scales [114]. The differences between the two scale profiles are small, but the sensitivity of the epitopes is enhanced by using three repeats from the domain center from the DNA binding domain [113]. Here it is the long-range stabilizing elastic interactions that enhance sensitivity, and not the short –range glycosylation ones [112], probably because glycosylation obstructs $\beta$ strand hydrogen bond interactions. The detailed analysis of the differences shown in Fig. 16 can be extended to other mucin epitopes studied in [113], with the conclusion that the largest scaling differences correlate well with largest diagnostic sensitivity differences [114].

The direct approach to treating cancer is supported by the Developmental Therapeutics Program (DTP) of the NCI, which maintains the infrastructure and expertise for the operation of cell-free and cell-based high-, medium- and low-throughput assays. The DTP functional genomics laboratory provides molecular analyses including gene expression microarrays, exon arrays, microRNA arrays, multiplexing gene assays, plus others as tools to determine the role of selected genes in the mechanism(s) of drug action and cellular responses to stressors [115].

The direct approach to a problem that involves identifying the $> 5$ key mutations (from a pool of $> 10$ million) that have affected $> 20$ checkpoints in a way dependent nonlinearly on individual

DNA's (containing > 20 million nucleotides) is advantageous in one respect: it will keep researchers busy for an indefinite period (probably longer than individual careers). However, suppose we reverse the problem, having already identified a small number of epitopic signals in the "linear response" regime of ECD (not the nonlinear, and fully entangled regime of diagnosed tumors). Then, already "knowing the answer", we can work backwards through a data base of millions of patients, to find patterns in the individual DNA mutations that correlate with epitopic patterns (which could grow to include dozens of individual epitopes). That would be a new chapter in the history of science and medicine.

## 17. Schrodinger's Dream and Kauzmann's Insights

In his enormously popular Dublin 1943 lectures and book, "What Is Life?", Erwin Schrodinger proposed that we could progress in answering this question by using thermodynamics. A few years later, high energy physicists gave us hydrogen bombs, while solid state physicists gave us transistors and the structure of DNA. Over the next 50 years high energy physicists produced nuclear power and the LHC, while solid state physicists gave us solar power, the Internet, and billions of fantastic gadgets. Molecular biologists joined molecular solid state physicists to generate an enormous data base of protein structures and functions, which have formed a platform for miraculous medical treatments.

The historical path from Schrodinger's classical discussion to modern thermodynamic scaling involves many modern technical tools from mathematics and theories of phase transitions. It also benefitted from intuitive insights, and here special mention should be made of Walter Kauzmann's work in the 1950's [116]. Kauzmann emphasized that proteins function reversibly on long time scales (ms), and behave like a deeply supercooled liquid with very high viscosity. Evolution has achieved Kauzmann's network qualities by going beyond classical glassy "funnels" [26] and approaching fractal critical points. Kauzmann anticipated the modern viewpoint through his emphasis on the key role played by hydropathic forces in shaping protein globules. Thermodynamic scaling may enable us to realize Schrodinger's dream, and advance new medical platforms where earlier work has stalled. Specifically it exploits the protein data base to describe the connections between amino acid sequences and protein functions with the accuracy of Schrodinger's dream and the intuitive insights of Kauzmann.

*Postscript.*    Simple laboratory toy models of sub-critical self-organized in/out globular shaping exhibit a variety of long-range interactions between surface droplets, analogous to protein amino acid very long-range interactions [117].    There is much interest in Hub proteins, which reside on the cytoplasmic side of the cell membrane [118], also described as a catalytic substrate supporting protein-protein interactions in the interfacial frontier space [119]. Profiles of some of these 8 proteins (examples in their Table 2, lengths range from 304 aa to 976 aa) could be interesting.    The connection between intrinsic disorder, conformational dynamics and Newtonian simulations has been reviewed (574 references) [119].    NMR experiments have shown that conformational dynamics in the catalytically relevant microsecond to millisecond timescale is optimized along the favored evolutionary trajectory in a bacterial protein [120]. Statistical methods alone applied to genome intron/extron analysis failed to identify ordering of protein sequences [121].  This illustrates why the combination of topological, geometrical and structural data in scaling theory is needed to connect protein sequences to function and biomedical applications. Students will find a concise, self-contained introduction to many of the classical tools used here in [122].

J. C. Phillips is an applied theoretical physicist, who studies semiconductors (basic to the opto-electronic revolution), network glasses (Gorilla glass), and protein sequences (since 2007).  He is an NAS (USA) member.

# References

1.  Liu, JF, Rost, B  Comparing function and structure between entire proteomes.  Prot. Sci.  **10,** 1970-1979  (2001).

2.  Liberles DA, Teichmann SA, Bahar I, Bastolla U, Bloom J, et al. The interface of protein structure, protein biophysics, and molecular evolution.  Prot. Sci. **21**, 769-785 (2012).

3.  Maragakis P, Karplus M Large amplitude conformational change in proteins explored with a plastic network model: Adenylate kinase. J. Mol. Biol. **352,** 807-822 (2005).

4.  Brooks CL, Karplus M Solvent effects on protein motion and protein effects on solvent motion - dynamics of the active-site region of lysozyme.  J. Mol. Biol. **208**, 159-181 (1989).

5.  Phillips JC Scaling and self-organized criticality in proteins: Lysozyme c. Phys. Rev. E **80**, 051916 (2009).

6.  Wallace ML, Lariviere V, and Gingras Y Modeling a century of citation distributions.  J. Informetrics, **3**, 296-303 (2009).

7.  Naumis GG,  Phillips  J C  Diffusion of knowledge and globalization in the Web of twentieth century science.  Phys. A-Stat. Mech. Appl. **391**, 3995-4003 (2012).

8.  Phillips JC Phase transitions in the web of science.  Phys. A **428**, 173-177 (2015).

**9.**  Symonds N What is Life?: Schrodinger's influence on biology.  Quart. Review Biolo. **61**, 221 (1986).

**10.** Holliday R  Physics and the origins of molecular biology. J. Genetics **85**, 93-97 (2006).

**11.** Cipra BA An introduction to the Ising model. Am. Math. Monthly **94**, 937-959 (1987).

12.  Wilson KG, Kogut J The renormalization group and the epsilon expansion. Phys. Rep. **12c**, 75-200 (1974).

13. Mandelbrot BB The fractal geometry of nature (Freeman, 1982).

14. Bak P,  Tang C, Wiesenfeld K Self-organized criticality: an explantion of 1/f noise. Phys. Rev. Lett. **59**, 381-384 (1987).

15. Bak P, Sneppen, K Punctuated equilibrium and criticality in a simple-model of evolution. Phys. Rev. Lett. **71**, 4083-4086(1993).

16. Marsh JA, Teichmann S A Parallel dynamics and evolution: Protein conformational fluctuations and assembly reflect evolutionary changes in sequence and structure. Bioessays **36**, 209-218 (2014).

17. Jack B R, Meyer AG, Echave J, et al. Functional sites induce long-range evolutionary constraints in enzymes. PLOS Bio. **14** , e1002452 (2016).

18. Brunori M, Gianni S Molecular medicine - To be or not to be. .Biophys. Chem. **214**, 33-46 (2016).

19. Chiti F, Stefani M, Taddei N, et al. Rationalization of the effects of mutations on peptide and protein aggregation rates. Nature **424**, 805-808 (2003).

20. Palliser CC, Parry DAD Quantitative comparison of the ability of hydropathy scales to recognize surface beta-strands in proteins. Prot. Struc. Func. Gen. **42**, 243-255 (2001).

21. Manavalan P, Ponnuswamy PK (1978) Hydrophobic character of amino acid residues in globular proteins. Nature **275**, 673–674.

22. Rose GD, Geselowitz AR, Lesser GJ, Lee RH, Zehfus MH  (1985) Hydrophobicity of amino acid residues in globular proteins. Science **229**, 834–838.

23. Soyer A, Chomilie, J, Mornon JP, et al. Voronoi tessellation reveals the condensed matter character of folded proteins.  Phys. Rev. Lett. **85**, 3532-3535 (2000).

24. Morais PA, Oliveira EA, Araujo NAM, et al. Fractality of eroded coastlines of correlated landscapes. Phys. Rev. E **84**, 016102 (2011).

25. Moret M A, Zebende G F Amino acid hydrophobicity and accessible surface area. Phys. Rev. E **75**, 011920 (2007).

26. Dill KA, Chan HS (1997) From Levinthal to pathways to funnels. Nature Struc. Biol. **4**, 10-19.

27. Moore AD, Bjorklund, AK, Ekrnan D, et al. Arrangements in the modular evolution of proteins. Trends Biochem. Sci. **33**, 444-451 (2008).

28. Machta B B, Chachra R, Transtrum M. K, Parameter Space Compression Underlies Emergent Theories and Predictive Models. Sci. **342**, 604-607 (2013).

29. Kyte J, Doolittle RF A simple method for displaying the hydropathic character of a protein. J. Mol. Biol. **157,** 105-132 (1982).

30. Fujiwara K, Toda H, Ikeguchi M Dependence of alpha-helical and beta-sheet amino acid propensities on the overall protein fold type. BMC Struc.Biol. **12**, 18 (2012).

31. de Groot NS, Pallares I,. Aviles FX et al. Prediction of "hot spots" of aggregation in disease-linked polypeptides. BMC Struc. Biol. **5**, 18 (2005).

32. Karmakar S, Dasgupta C, Sastry S Length scales in glass-forming liquids and related systems: a review. Rep. Prog. Phys. **79**, 016601 (2016).

33. Lemaitre A Structural relaxation is a scale-free process. Phys. Rev. Lett. **113**, 245702 (2014).

34. Thorpe M.F, Jacobs D J, Chubynsky MV, et al Self-Organization in Network Glasses. J. Non_Cryst. Sol. **266-269**, 859 (2000).

35. Micoulaut M Relaxation and physical aging in network glasses. Rep. Prog. Phys. **79**, 06654 (2016).

36. Mauro JC, Tandia A, Vargheese KD, et al. Accelerating the design of functional glass through modeling. Chem. Mater. **28**, 4267−4277 (2016).

37. Sinha N, Kumar S, Nussinov R (2001) Interdomain interactions in hinge-bending transitions. Structure **9**, 1165-1181.

38. Yang L, Song G, Jernigan RL (2007) How well can we understand large-scale protein motions using normal modes of elastic network models? Biophys. J. **93**, 920-929.

39. Cardamone L, Laio A, Torre V, Shahapure R, DeSimone A Cytoskeletal actin networks in motile cells are critically self-organized systems synchronized by mechanical interactions. Proc. Nat. Acad. Sci. (USA) **108**, 13978-13983 (2011).

40. Yang L, Song G, Jernigan RL Protein elastic network models and the ranges of cooperativity. Proc. Nat. Acad. Sci. (USA) **106**, 12347-12352 (2009).

41. Geeves MA, Holmes KC Structural mechanism of muscle contraction. Ann. Rev. Biochem. **68**, 687-728 (1999).

42. Yildiz A, Tomishige M, Vale RD, et al. Kinesin walks hand-over-hand. Science **303**, 676-678 (2004).

43. Cardamone L, Laio A, Torre V, Shahapure R, DeSimone A (2011) Cytoskeletal actin networks in motile cells are critically self-organized systems synchronized by mechanical interactions. Proc. Nat. Acad. Sci. (USA) **108**, 13978-13983.

44. Goyon, J.; Colin, A.; Ovarlez, G.; et al. Spatial cooperativity in soft glassy flows. Nature 454, 84-87 (2008).

45. Osher S, and Sethian JA Fronts Propagating with Curvature-Dependent Speed: Algorithms Based on Hamilton--Jacobi Formulations. J. Comp. Phys. **79**, 12—49 (1988).

46. Malladi R, Sethian JA, Vemuri BC Shape modelling with front propagation – a level set approach. IEEE Trans. Patt. Anal. Mach. Intel. 17, 158-175 (1995).

47. Chan, TF; Vese, LA Active contours without edges. IEEE Trans. Image Proc. **10**, 266-277 (2001).

48. Saye RI, Sethian JA The Voronoi Implicit Interface Method for computing multiphase physics. Proc. Nat. Acad. Sci. (USA) **108**, 19498-19503 (2011).

49. Manavalan P, Ponnuswamy PK Hydrophobic character of amino acid residues in globular proteins. Nature **275**, 673–674 (1978).

50. Phillips JC Fractals and Self-Organized Criticality in Proteins. Phys. A **415**, 440-448 (2014).

51. Mountain RD, Spectral distribution of scattered light in a simple fluid. Rev. Mod. Phys. **38**, 205-214 (1966).

52. Y. Sugimoto, Y. Kamada, Y. Tokunaga, et al. Aggregates with lysozyme and ovalbumin show features of amyloid-like fibrils. Biochem. Cell Biol. **89**, 533-544 (2011).

53. Y. Tokunaga,; M. Matsumoto, Y. Sugimoto, Amyloid fibril formation from a 9 amino acid peptide, 55th-63rd residues of human lysozyme. Int, J. Bio. Macromol. **80**, 208-216 (2015).

54. Cardamone L, Laio A, Torre V, et al. Cytoskeletalactin networks in motile cells are critically self-organized systems synchronized by mechanical interactions. Proc. Nat. Acad. Sci. (USA) **108**, 13978-13983 (2011).

55. Hatzakis NS Single molecule insights on conformational selection and induced fit mechanism. Biophys. Chem. **186**, 46-54 (2014).

56. Schiller JT, Lowy DR Understanding and learning from the success of prophylactic human papillomavirus vaccines. Nature Rev. Microbio. **10**, 681-692 (2012).

57. Phillips JC Minimal immunogenic epitopes have nine amino acids. arXiv (2016).

58. Jourdier S A miracle drug. Chemistry in Britain **35**, 33-35 (1999).

59. Vane JR, Botting RM The mechanism of action of aspirin. Throm. Res. **110**, 255-258 (2003).

60. Smith WL,.DeWitt DL, Garavito RM Cyclooxygenases: Structural, cellular, and molecular biology. Ann. Rev. Biochem. **69**, 145-182 (2000).

61. Rouzer CA, Marnett L J (2009) Cyclooxygenases: Structural and functional insights. J. Lipid Res. **50**, S29-S34.

62. Luong C, Miller A, Barnett J, et al. (1996) Flexibility of the NSAID binding site in the structure of human cyclooxygenase-2. Nature Struc. Biol. **3**, 927-933.

63. Phillips, JC Fractals and Self-Organized Criticality in Anti-Inflammatory Drugs. Physica A **415**, 538-543 (2014).

64. Phillips, JC Hidden Thermodynamic Information in Protein Amino Acid Mutation Tables. arXiv 1601.03037 (2016).

65. Ordway GA, Garry, DJ Myoglobin: an essential hemoprotein in striated muscle> J. Exp. Biol. 207, 3441-3446 (2004).

66. Burmester T, Hankeln T (2009). What is the function of neuroglobin? J. Exp. Biol. **212**, 1423-1428.

67. Salter M D, Blouin G C, Soman J, et al. Determination of ligand pathways in globins: apolar tunnels versus polar gates.. J. Biol. Chem. **287**, 33163-33178 (2012).

68. Yu T-Q, Lapelosa M, Vanden-Eijnden, E, et al. Full kinetics of CO entry, internal diffusion, and exit in myoglobin from transition-path theory simulations. J. Am. Chem. Soc. **137**, 3041-3050 (2015).

69. Sachdeva V, Phillips, JC Oxygen Channels and Fractal Wave-Particle Duality in the Evolution of Myoglobin and Neuroglobin.

70. Ollis DL, Cheah E, Cygler DM, et al. (1992) The alpha/beta hydrolase fold. Prot. Eng. **5**, 197-211.

71. Buller AR, Townsend CA (2013) Intrinsic evolutionary constraints on protease structure, enzyme acylation, and the identity of the catalytic triad. Proc. Nat. Acad. Sci. (USA) **110**, E653-E661 (2013).

72. Phillips, JC Hidden Thermodynamic Information in Protein Amino Acid Mutation Tables.

73. Sachdeva V, Phillips, JC Hemoglobin Strain Field Waves and Allometric Functionality.

74. Yuan Y, Tam MF, Simplaceanu V, et al. New look at hemoglobin allostery. Chem. Rev. **115**, 1702-1724 (2015).

75. Phillips, JC Ineluctable Complexity of High Temperature Superconductivity Elucidated. J. Supercond. Novel Mag. **27**, 345-347 (2014).

76. Phillips JC Interference between resonance and potential scattering in UV spectra of insulators. Phys. Rev. Lett. **12**, 447-449(1964).

77. Ignatchenko VA, Polukhin DS Crossing resonance of wave fields in a medium with an inhomogeneous coupling parameter. J. Exp. Theor. Phys. **117**, 846-861 (2013).

78. Phillips JC Self-organized criticality and color vision: A guide to water-protein landscape evolution. Phys. A 392, 468-473 (2013).

79. Phillips JC Hydropathic self-organized criticality: a magic wand for protein physics. Protein & Peptide Lett. **19**, 1089-1093 (2012).

80. Mustafi D, Engel AH, Palczewski K (2009) Structure of cone photoreceptors. *Prog. Retin. Eye Res.* 28: 289-302.

81. Hill B, Roger T, Vorhagen FW (1997) Comparative analysis of the quantization of color spaces on the basis of the CIELAB color-difference formula. *ACM Trans. Graph.* 16: 109-154.

82. Lee BB, Martin PR, Grunert U (2010) Retinal connectivity and primate vision. *Prog. Ret. Eye Res.* 29: 622-639.

83. Naumis GG Phillips JC (2012) Bifurcation of stretched exponential relaxation in microscopically homogeneous glasses: *J Non-Cryst. Sol*. 358: 893-897.

84. Mazziotti DA *Phys. Rev. Lett.* **101** 253002 (2008).

85. Fujimoto K, Hasegawa J, Nakatsuji H (2009) Color tuning mechanism of human red, green, and blue cone pigments: SAC-CI Theoretical Study. *Bull. Chem. Soc. Jap.* 82: 1140-1148.

86. Millman D, Mihalas S, Kirkwood, A, et al. Self-organized criticality occurs in non-conservative neuronal networks during 'up' states. Nature Phys. **6**, 801- 805 (2010).

87. Camilloni C, Sala BM, Sormanni P, et al. Rational design of mutations that change the aggregation rate of a protein while maintaining its native structure and stability. Sci. Rep. **6**, 25559-25559 (2016).

88. Wong J, Sridharan S, Berdeprado J, et al. Predicting residual kidney function in hemodialysis patients using serum beta-trace protein and beta 2-microglobulin. Kid. Int. **89**, 1090-1098 (2016).

89. Phillips JC arXiv 1553532 (2015).

90. Drake JW, Holland JJ Mutation rates among RNA viruses. Proc. Natl. Acad. Sci. USA **96**, 13910-13913 (1999).

91. Poon LLM, Song T, Rosenfeld R, et al. Quantifying influenza virus diversity and transmission in humans. Nature Gen. **48**, 195-200 (2016).

92. Bush RM, Fitch WM, Bender CA, et al Positive selection on the H3 hemagglutinin gene of human influenza virus A. Mol. Biol. Evolu. **16**, 1457-1465 (1999).

93. Chan JM, Carlsson G, Rabadan R Topology of viral evolution.  Proc. Natl. Acad. Sci. USA **110**, 18566-18571 (2013).

94. Greenbaum BD, Ghedin E Viral evolution: beyond drift and shift.  Curr.  Opin. Microbio. **26**, 109-115 (2015).

95. Phillips JC Vaccine escape in 2013-4 and the hydropathic evolution of glycoproteins of A/H3N2 viruses. Physica A **452**, 38-43 (2016).

96. Phillips JC Punctuated evolution of influenza virus neuraminidase (A/H1N1) under migration and vaccination pressures.  BioMed Research Int. **2014**, 907381 (2014).

97. Earp LJ, Delos SE, Park HE, et al. The many mechanisms of viral membrane fusion proteins. Curr. Top. Microbio. Imm.  **285**, 25-66 (2005).

98. Phillips JC  Proteinquakes in the Evolution of Influenza Virus Hemagglutinin (A/H1N1) under Opposing Migration and Vaccination Pressures. BioMed Research Int.  **2015** 243162 (2015).

99. Gupta V, Earl DJ, Deem MW Quantifying influenza vaccine efficacy and antigenic distance.  Vaccine **24**, 3881-3888 (2006).

100. He J,  Deem MW Low-dimensional clustering detects incipient dominant influenza strain clusters.  Prot. Eng. Design, Selec. **23**, 935-946 (2010).

101. Earl DJ, Deem MW Parallel tempering: Theory, applications, and new perspectives. Phys. Chem. Chem. Phys. **7**, 3910-3916 (2005).

102. Li X, Phillips JC (unpublished)

103. Fujiwara K, Toda H, Ikeguchi M,  Dependence of alpha-helical and beta-strand amino acid propensities on the overall protein fold type. BMC Struc. Biol. **12**, 18 (2012).

104. de Groot N S, Pallares I, Aviles F X, et al. Prediction of "hot spots" of aggregation in disease-linked polypeptides.  BMC  Struc. Biol. **5**, 18 (2005).

105. Phillips JC  Catalytic nucleation of amyloid beta, hen egg white fibrils,  p53 oligomerization, and β-2 microglobulin.

106. DeVita VT, DeVita-Raeburn E, The Death of Cancer. Farrar, Strauss and Giroux, New York, p.287 (2015).

107.  Topalian SL, Drake CG, Pardol, DM Immune Checkpoint Blockade: A Common Denominator Approach to Cancer Therapy. Cancer Cell **27**, 450-461 (2015).

108. Pedersen JW, Gentry-Maharaj A, Fourkala E-O, et al. Early detection of cancer in the general population: a blinded case-control study of p53 autoantibodies in colorectal cancer. Brit. J. Cancer **108**, 107-114 (2013).

109. Mock A, Warta R, Geisenberger C, et al. Printed peptide arrays identify prognostic TNC serum antibodies in glioblastoma patients. Oncotarget **6**, 13579-13590 (2015).

110. Phillips JC Autoantibody recognition mechanisms of p53 epitopes. Physica A **451**, 162-170 (2016).

111. Goodsell DS The molecular perspective: p53 tumor suppressor. The oncologist **4**, 138-139 (1999).

112. Pedersen JW, Blixt O, Bennett EP, et al. Seromic profiling of colorectal cancer patients with novel glycopeptide microarray. Int. J. Cancer **128**, 1860-1871 (2011).

113. Pedersen JW, Gentry-Maharaj A, Nostdal, A, et al. Cancer-associated autoantibodies to MUC1 and MUC4 - A blinded case-control study of colorectal cancer in UK collaborative trial of ovarian cancer screening. Int. J. Cancer **134**, 2180-2188 (2014).

114. Phillips JC Autoantibody recognition mechanisms of MUC1. arXiv 1606.07024 (2016).

115. Teicher BA Perspective: Opportunities in recalcitrant, rare and neglected tumors. Oncol. Rep. 30, 1030-1034 (2013).

116. Kauzmann W, Some Factors in the Interpretation of Protein Denaturation. Adv. Prot. Chem. **14**, 1–63. (1959).

117. Jagota A Role reversal: Liquid "Cheerios" on a solid sense each other. Proc. Nat. Acad. Sci. (USA) **113**, 7294-7295 (2016).

118. Ota M, Gonja H, Koike R, et al. Multiple-Localization and Hub Proteins. PLOS ONE **11**, e0156455 (2016).

119. Wei G, Xi W, Nussinov R, et al. Protein Ensembles: How Does Nature Harness Thermodynamic Fluctuations for Life? The Diverse Functional Roles of Conformational Ensembles in the Cell. Chem. Rev. **116**, 6516-6551 (2016).

120. Gonzalez M, Abriata LA, Tomatis PE, et al. Optimization of Conformational Dynamics in an Epistatic Evolutionary Trajectory. Mol. Bio. Evol. **33**, 1768-1776 (2016).

121. Guharay S, Hunt BR, Yorke JA, et al. Correlations in DNA sequences across the three domains of life. Phys. D **146**, 388-396 (2000).

122. Peliti L Statistical Mechanics in a Nutshell. Princeton Univ. Press (2011).

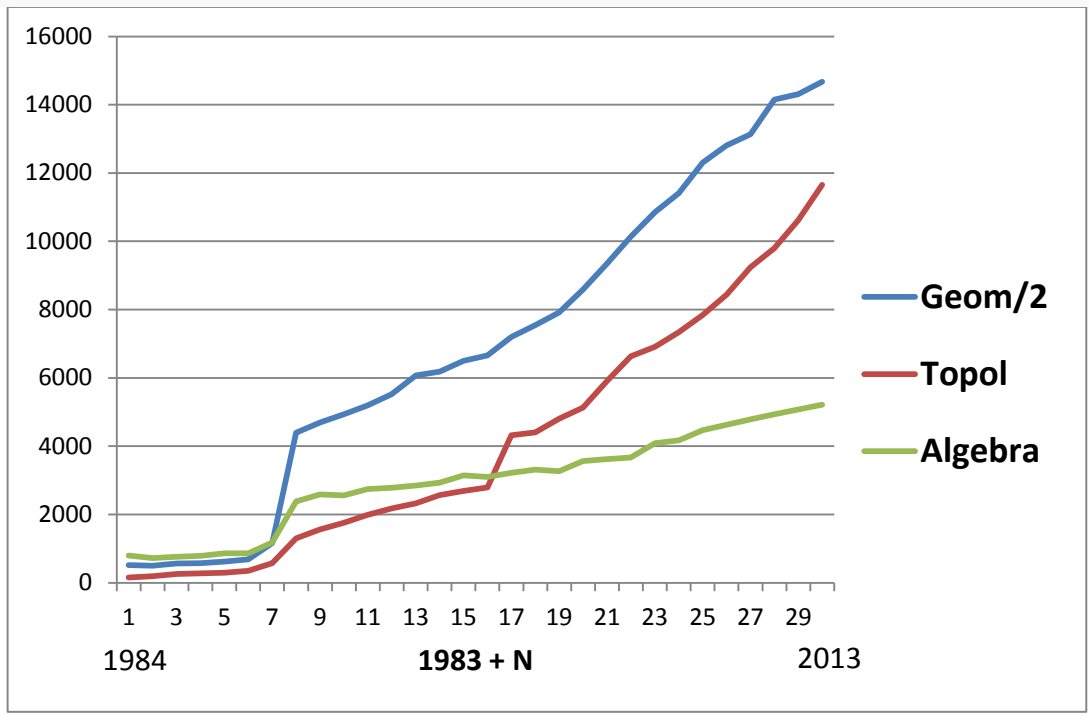

Fig. 1  The annual numbers of papers on each of three mathematical disciplines, with the geometry number divided by 2.  The entire geometry profile covers 330,000 papers.  Note not only the abrupt increases around 1990 associated with glasnost, but also the rapid increase in topology, which crosses over algebra with the advent of the Internet in 2000.

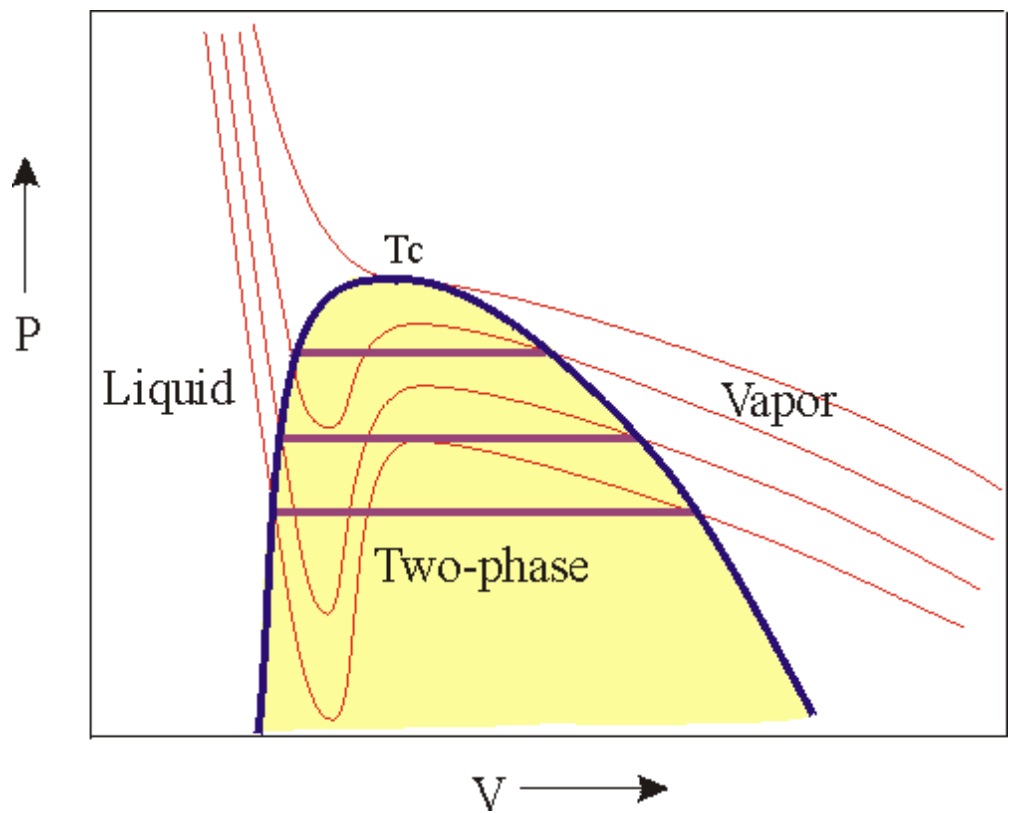

Fig. 2. Cubic isothermal curves for the van der Waals equation of state. For T > T$_c$, the curves are monotonic and contain no extrema. For T < T$_c$, there are two extrema, and phase separation occurs, with the liquid and vapor phases connected by the purple tie lines. The two extrema merge at the critical temperature T$_c$. From R. L. Rowley, Web Module: Van der Waals Equation of State, with author's permission.

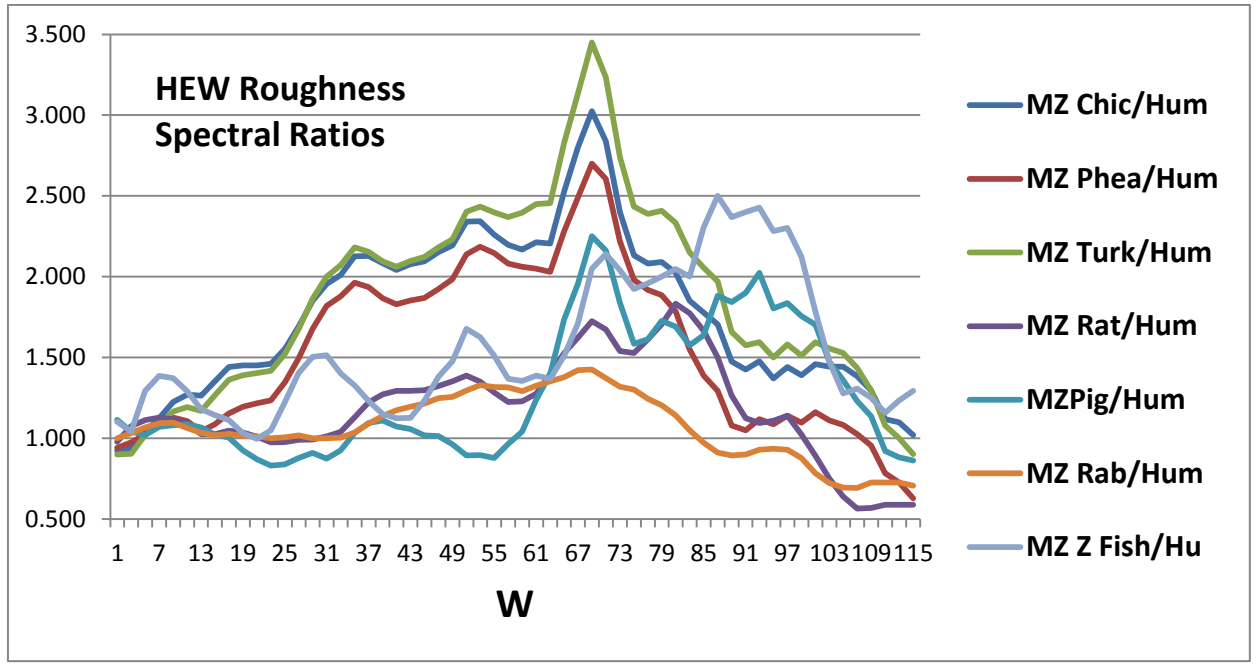

Fig. 3. The roughness $R_S(W)/R_{Human}(W)$ ratios are shown for six terrestrial species [8,50]. These ratios exhibit both a broad and narrow peak, both centered on W = 69, suggesting that the evolution of HEW from chicken to human has been directed towards improving a specific function, avoiding aggregation (amyloidosis).

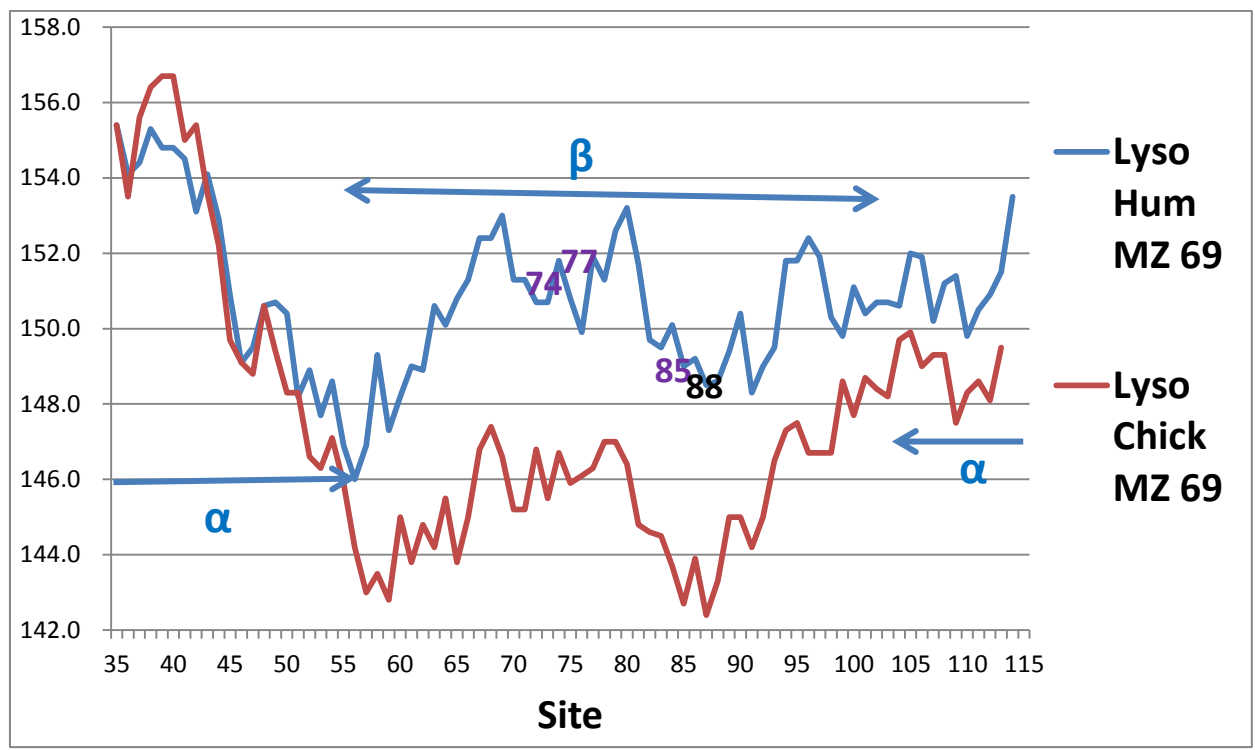

Fig. 4. Hydropathic profiles of human and chicken HEW, using the fractal MZ scale, with W = 69, as suggested by Fig. 3. Note the large stabilization by the human strain in the β region, compared to the small differences in the α regions. Here increasing ordinate corresponds to increasing hydrophobicity and increasing rigidity. The numbered sites near the center show large mutated aggregation rate changes.

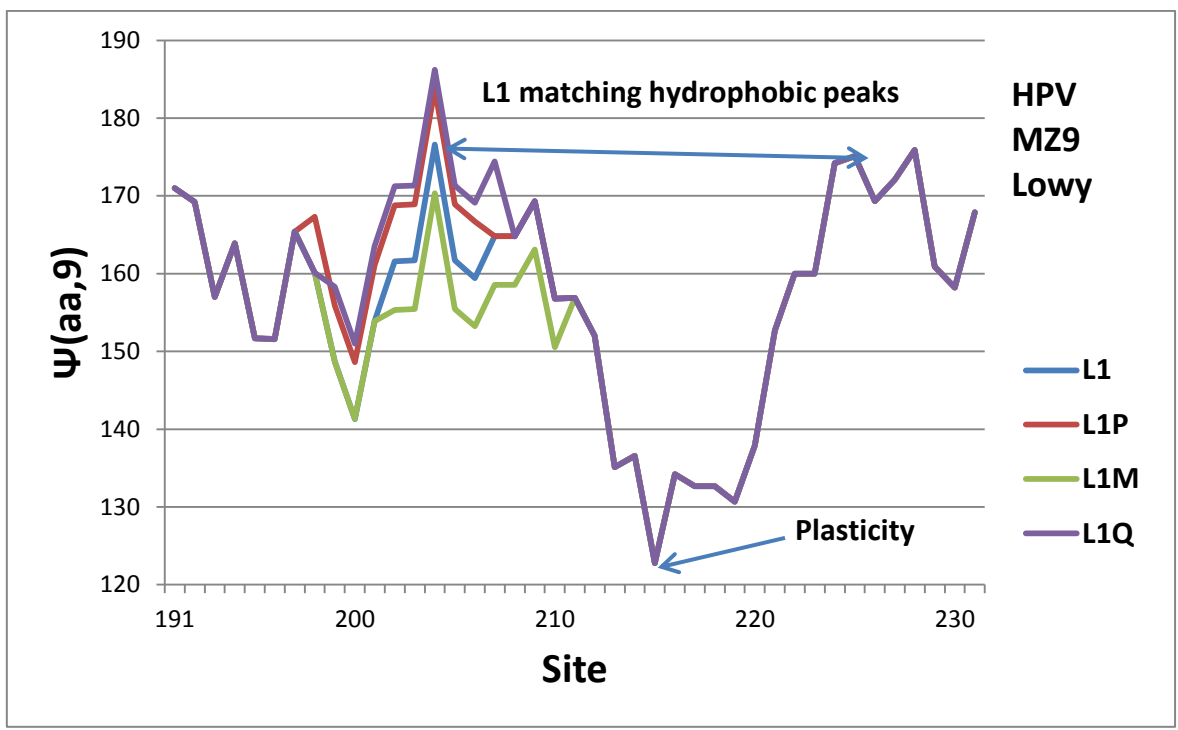

Fig. 5. Hydroprofile of L1 and several single amino acid mutants,, using the fractal scale [25,55].

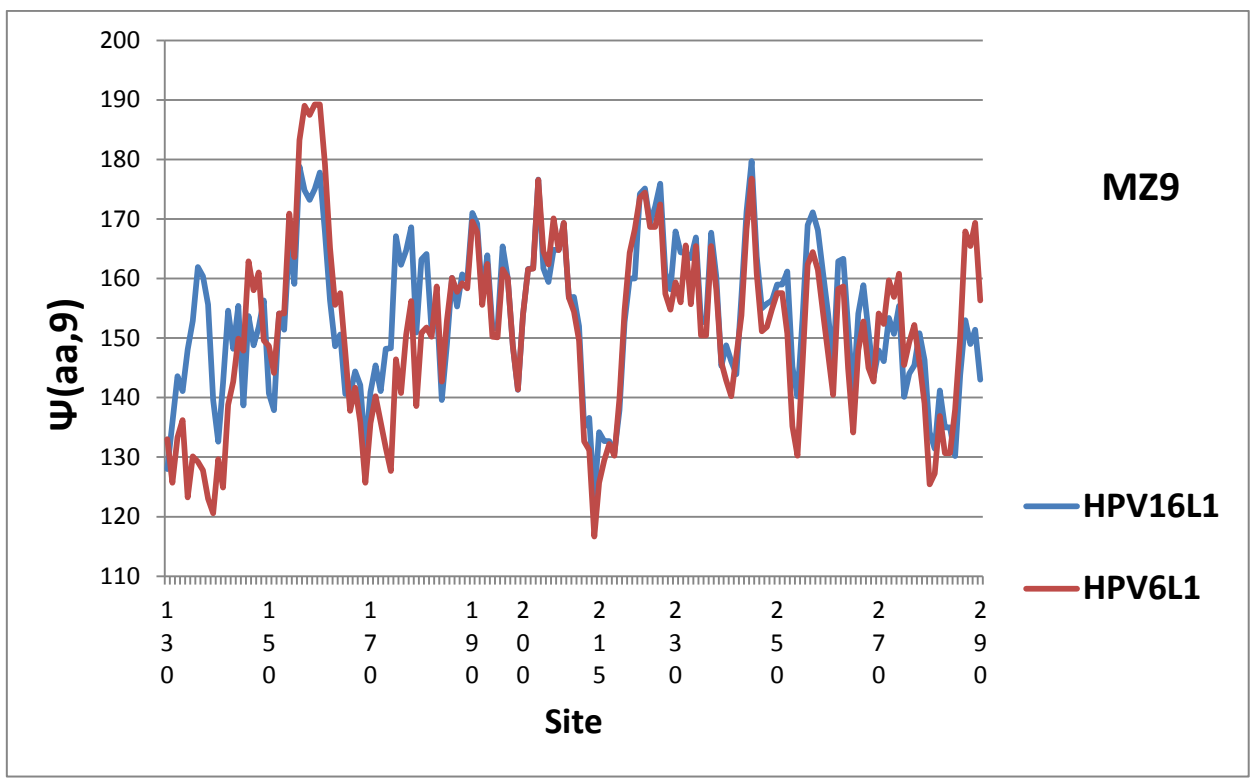

Fig. 6. By aligning the extrema, the MZ9 hydropathic profile reveals strong similarities (r = 0.82) between L1 HPV16 (cervical cancer) and HPV6 (warts).  The large differences around 135 and 160 could be the major factor in the functional differences.

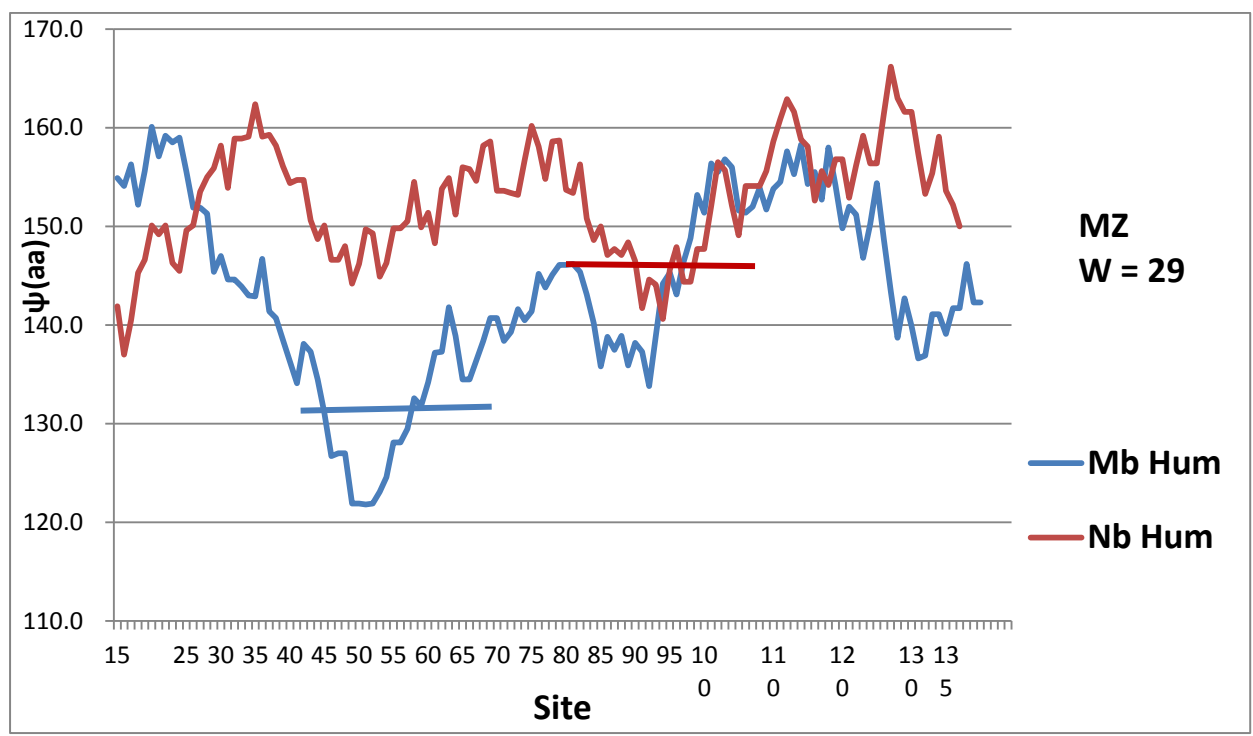

Fig. 7. The Ψ(aa,W= 29) hydropathic profiles for human Myoglobin and Neuroglobin. The ψ(aa) averages for the deepest elastic region minima spanning 29 amino acids are marked. The much lower elastic blue value for Mb enables the E8 His (64) gate to open and function as a small gas molecule channel. The much larger red elastic value for Ngb stiffens this region, so that the functional channel is switched from Mb. Note that in Mb the deepest hydrophilic minimum is centered near the distal His 64, while in Ngb the deepest minimum is centered near the proximate His 96. This shift is associated with the two different oxidation channels. Note also that the two Ngb minima are nearly level, suggesting a faster synchronized oxygen release.

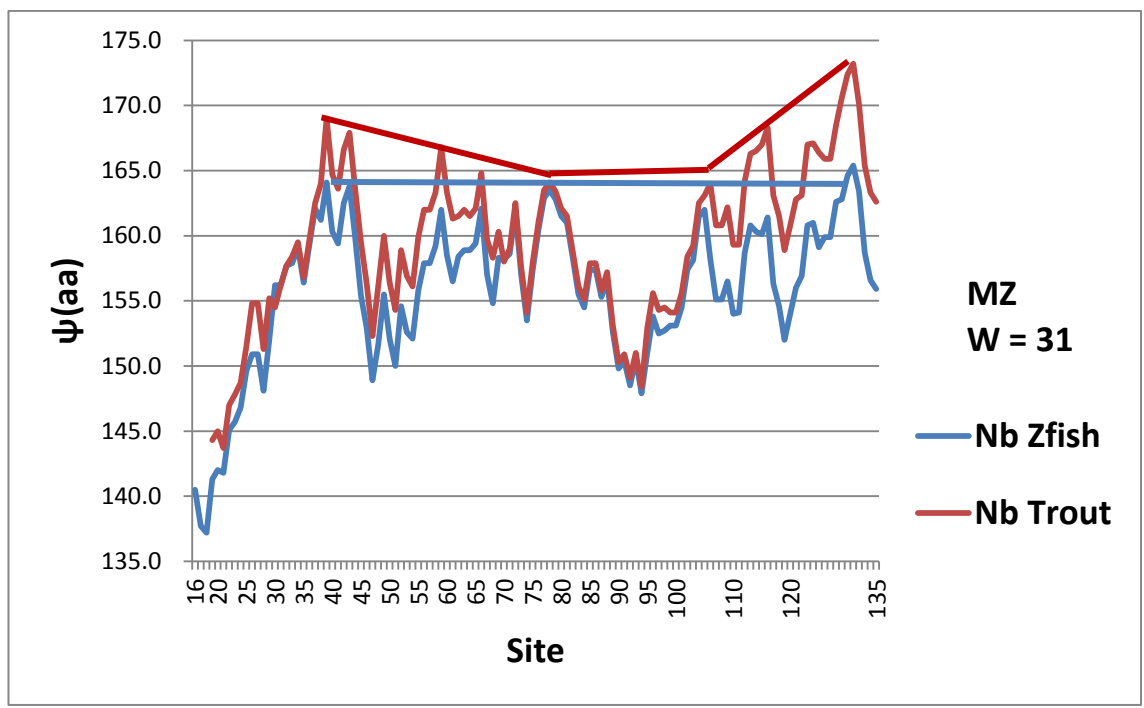

Fig. 8. Comparison of the hydroprofiles of tropical and temperate freshwater fish. The central region involving the heme and the distal His 64 – proximate His 96 channel is conserved, while the remaining regions are parabolically more hydrophobic for temperate trout, and nearly level for tropical Zebrafish. The parabolically more hydrophobic regions are necessary because of larger temperature fluctuations in temperate regions. Note also the unmarked hydrophilic minima at 47 and 94. These also divide the proteins into three nearly equal regions, with the Zebrafish hydrophilic minima again being more nearly level [69].

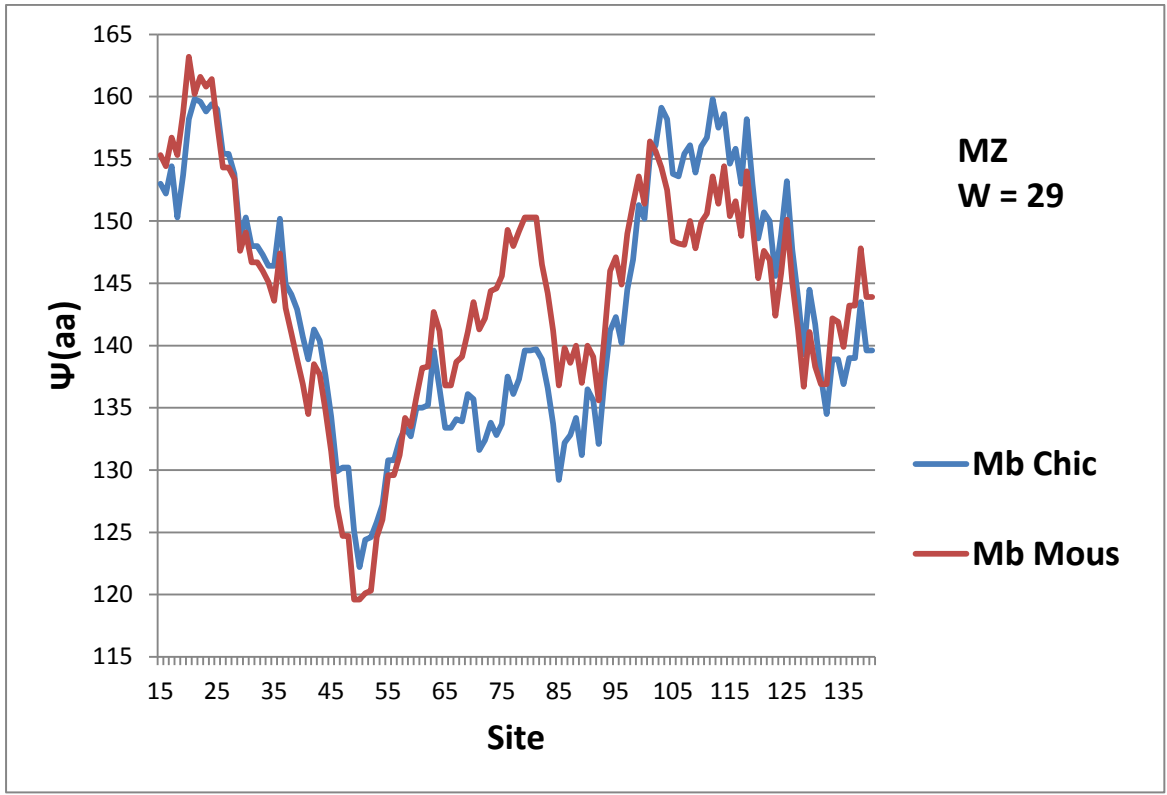

Fig. 9. The largest change here is the mouse apical hydroneutral peak near site 80. It acts as a balanced pivot, which facilitates release of both O and CO.

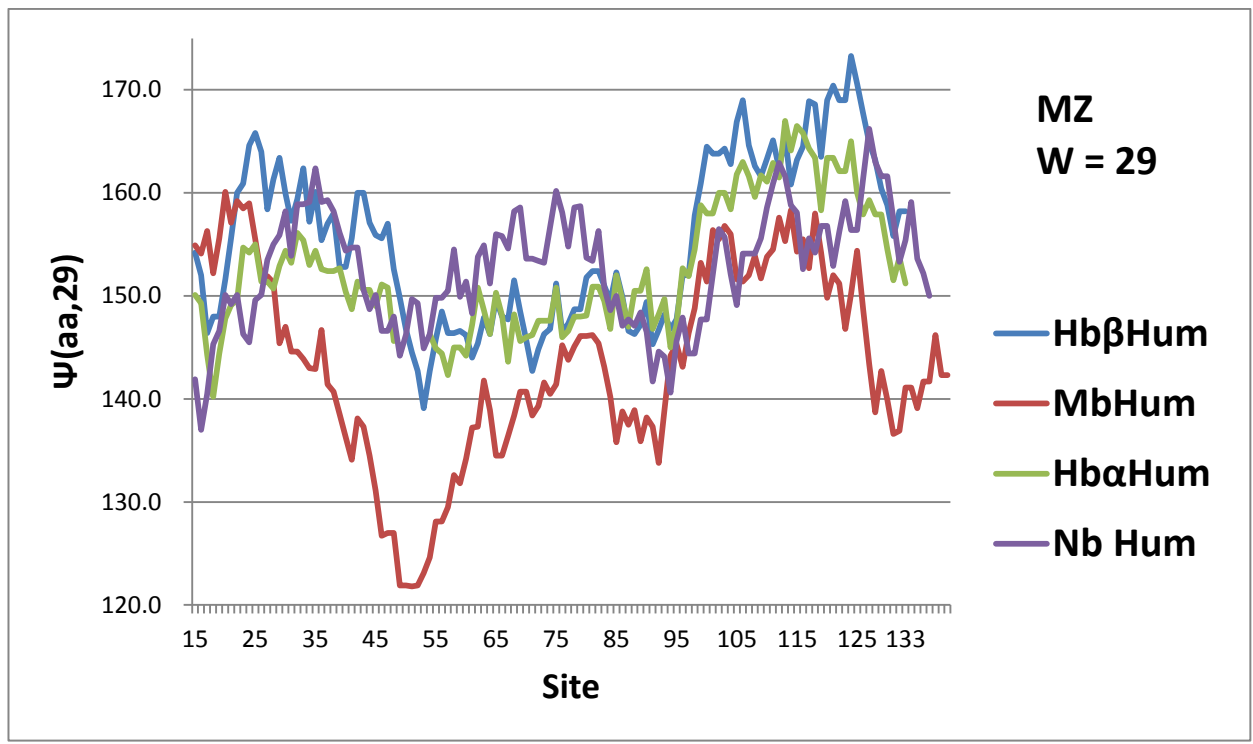

Fig. 10.   A striking feature of these profiles is that Mgb is the "odd globin out", which can be understood as reflecting the function of Mgb as storing oxygen in tissues for long periods.

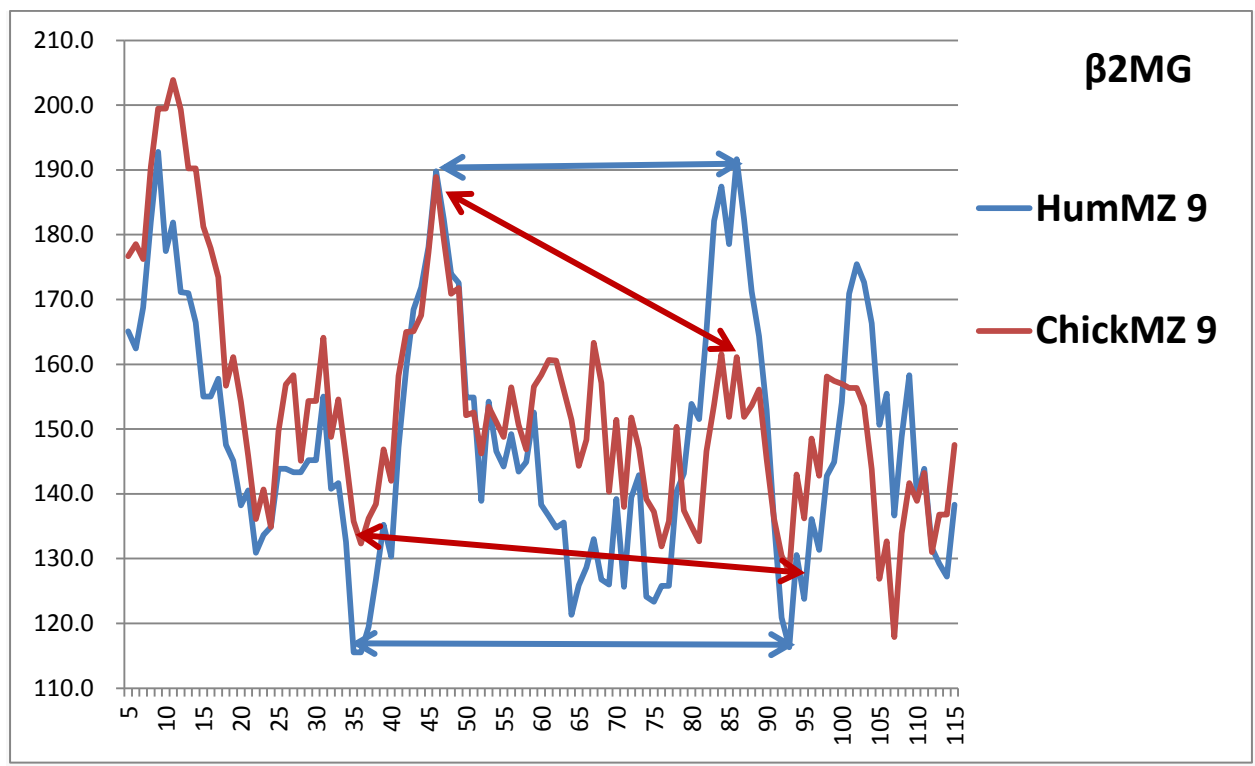

Fig. 11. Profiling the hydropathic shape of β2m with W = 9 reveals striking features, and how these have evolved. The structure is stabilized by two hydrophobic peaks near sites 45 and 85, while its kinetics are driven by the hydropathic amide ends and the hydrophilic carboxyl ends. There are two hinges near 35 and 93, which are nearly level in chicken, and almost exactly level in human. The largest evolutionary change is the leveling of the two hydrophobic peaks near sites 45 and 85 in humans.

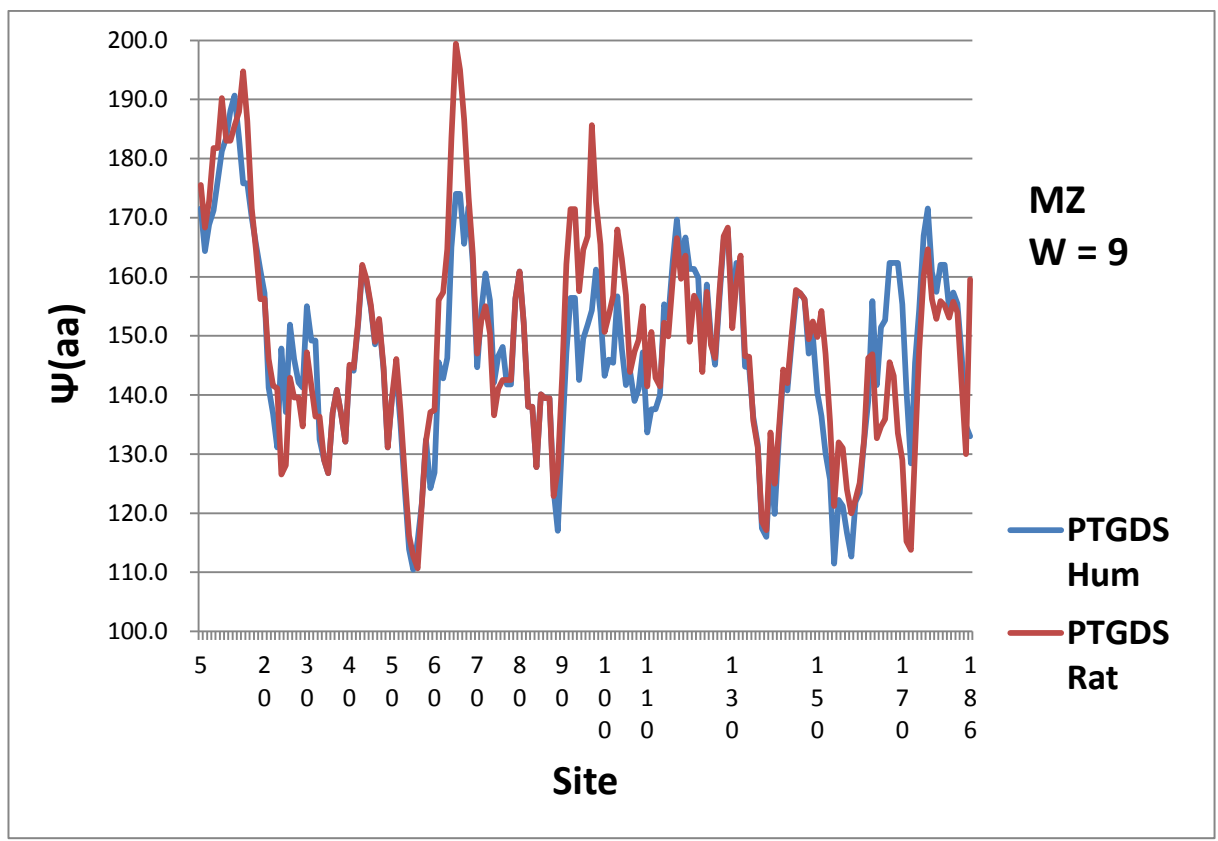

Fig. 12. Comparison of human and rat prostaglandin-H2 D-isomerase MZ W = 9 profiles. The human profile is smoother and flatter across multiple hydrophobic and hydrophilic extrema, which enable this enzyme to function better in multiple contexts [63].

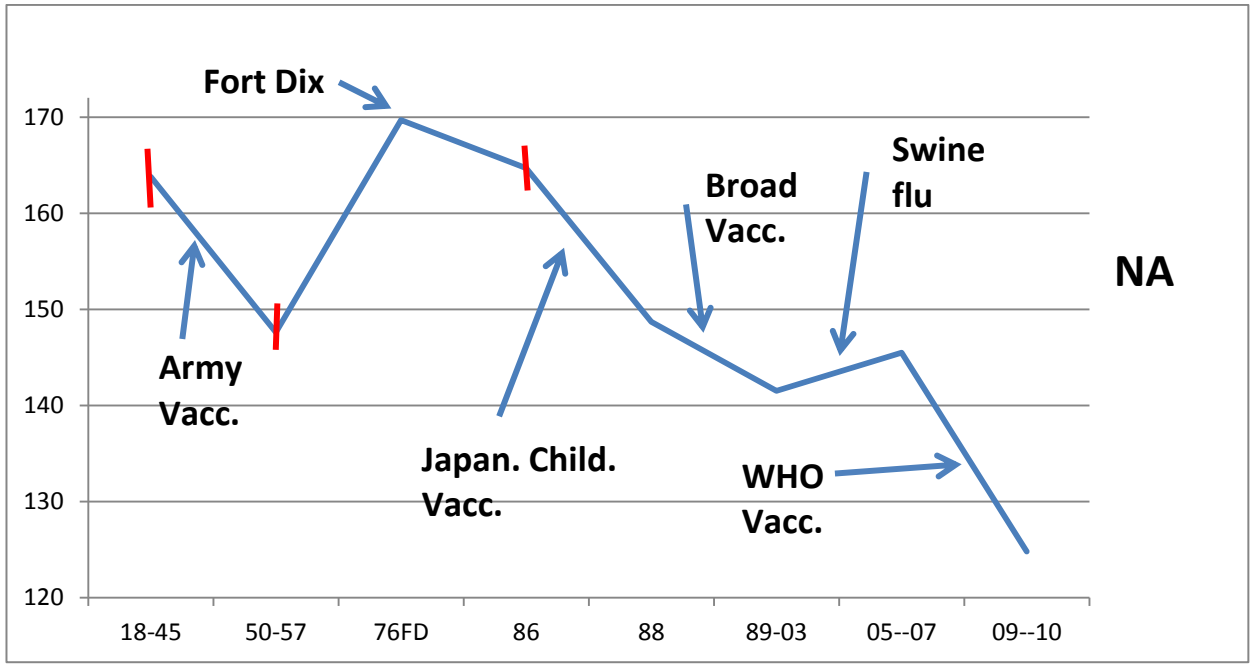

Fig. 13. A panoramic presentation of the opposing effects of migration and vaccination on NA roughness (variance of $\psi(aa,17)$, MZ scale) after the first wide-spread vaccination program, begun by the US Army in 1944, as listed in Table I. Flu virulence decreases or increases in tandem with NA roughness. The build-up of swine flu virulence from 2001 on is evident in selected urban areas (New York, Berlin, Houston) with crowded immigrant neighborhoods . A few early error bars are indicated in red, but after 1986 these become too small for this sketch.

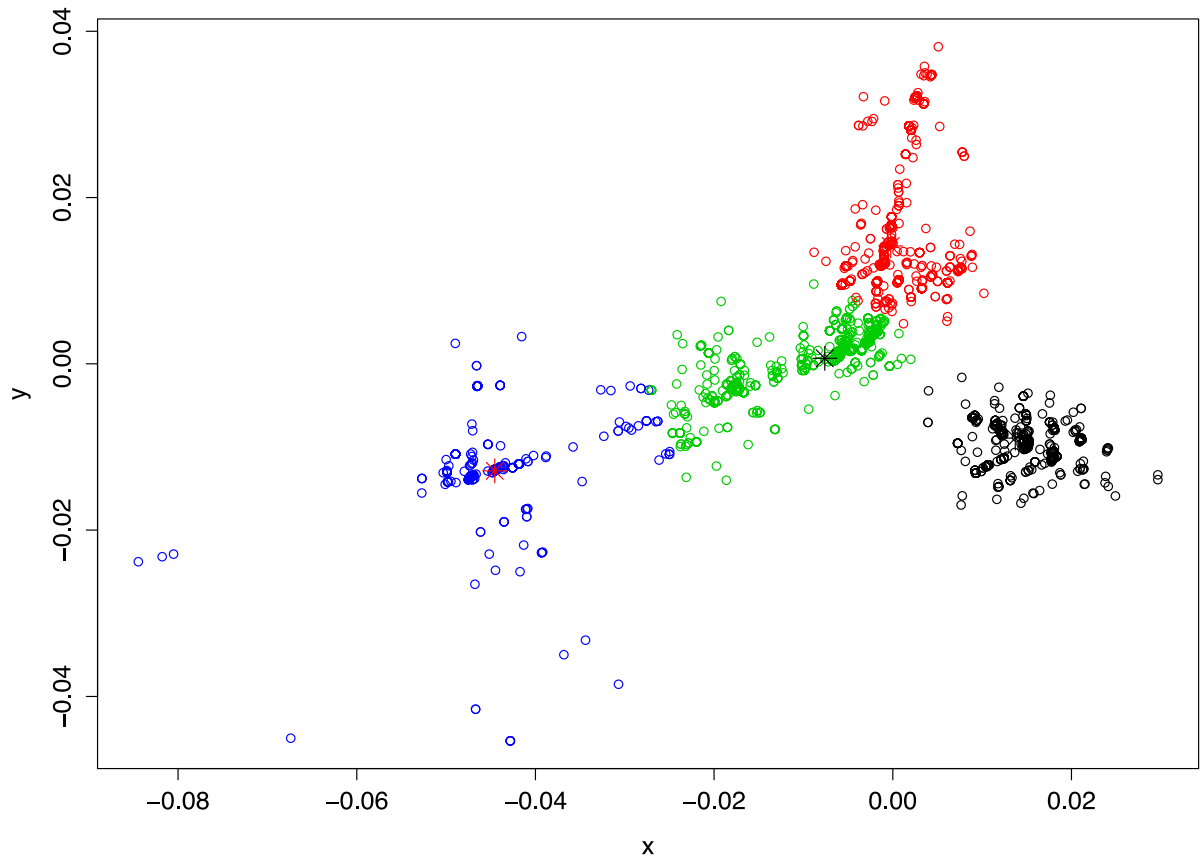

Fig. 14 Voronoi clustering of d = 2 compressed sequence data. Colors represent for 4 different clusters, each centering at the stars. The clusters follow a linear time orderliness from left to right. Black area is the A/Nebraska/4/2014 cluster. A curious and as yet unexplained feature of the three earlier clusters is their linear backbones. The time intervals for these four clusters is altogether 6 months (1.5 months each, 15.7.1-15.12.31), which illustrates how timely predictions based on it can be. Most phylogenic plots use at least 12 months per cluster, and have much weaker resolution.

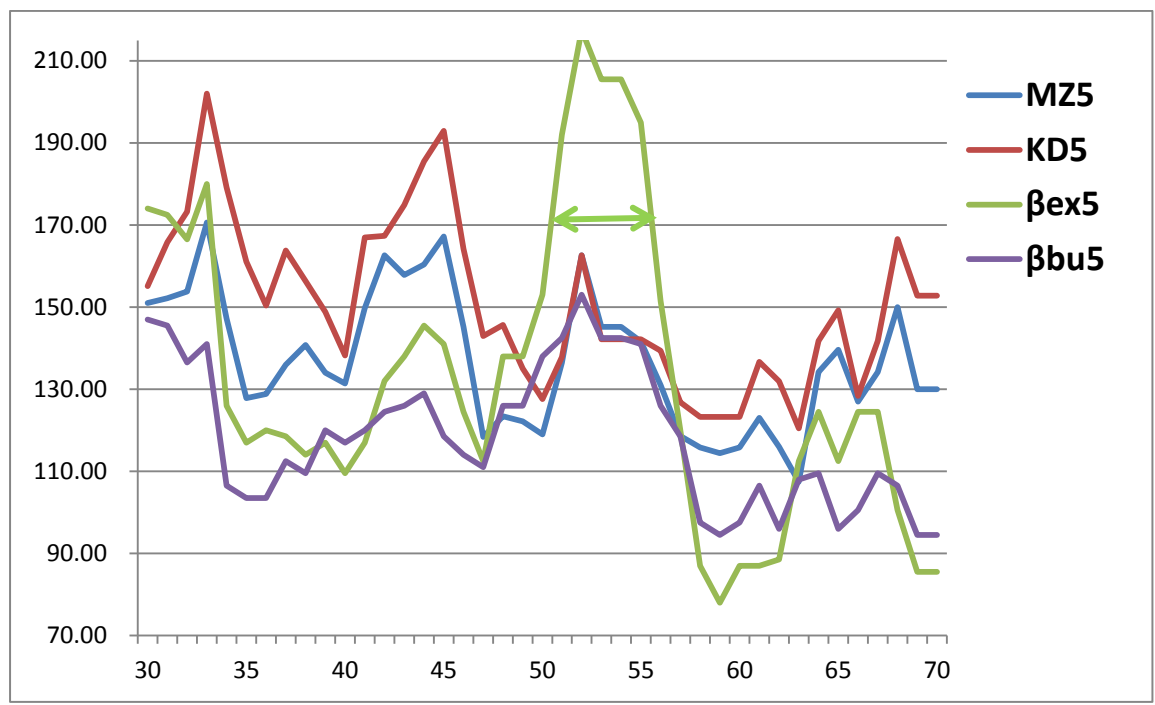

Fig. 15. [110] used four scales (two Ψ, two β) in the 30 − 70 site range where the sensitive epitopes -9-10 (sites 41-60) occur, with W = 5. The choice of a small W produces noisy profiles, but the success of the βex scale compared to the other scales in identifying the central 7-mer 50-56 IEQWFTE epitope is clear (double arrow). Note the Trp53 = W53 at the center of this epitope, and how use of the βex scale lifts the 50-56 compressed epitope signal well above noise level.

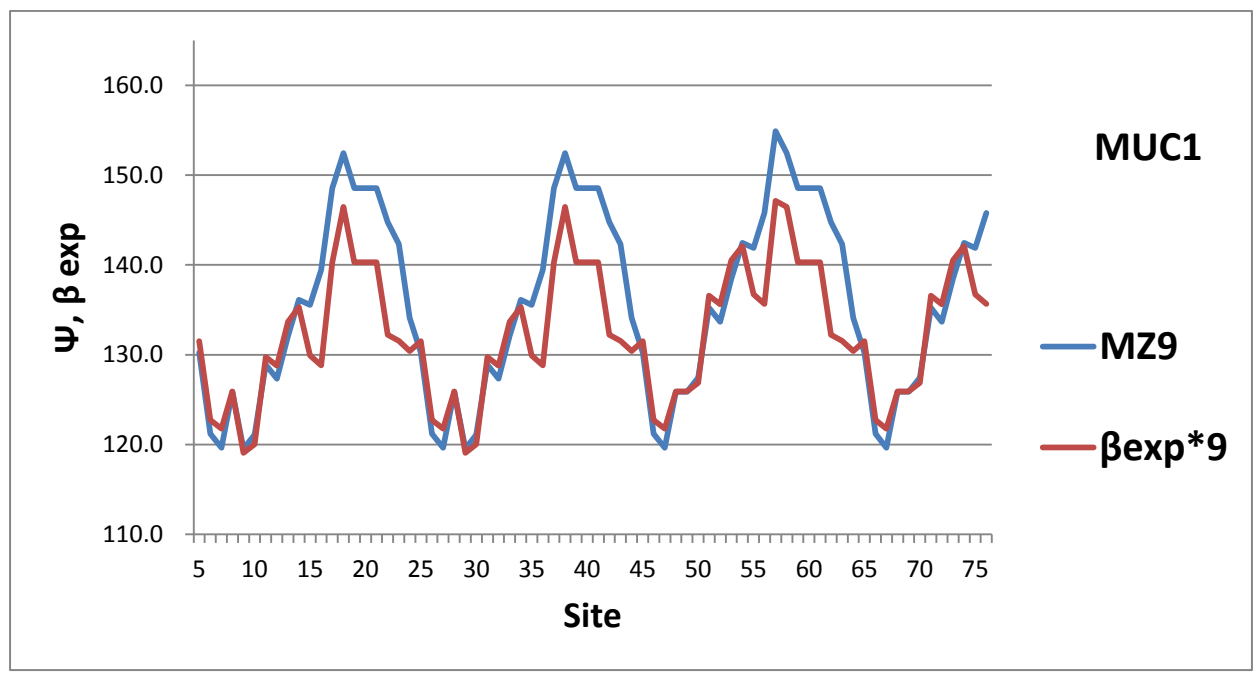

Fig. 16. For the most sensitive MUC1 repeats the βexp* values lie below the ΨMZ values (W = 9), but in a special way: about 10 near the relatively hydrophobic maxima, but near 0 for the extremely hydrophilic minima. Overall the MUC1 ΨMZ values are hydrophilic, and lie below the hydroneutral value for ΨMZ of 155. This is consistent with the overall disordered mucin structures, and is similar to the overlapping 15-mer 40-60 p53 region of Fig. 15.